\documentclass[aps,prd,reprint]{revtex4-2}

\usepackage[colorlinks]{hyperref}
\usepackage{amsmath}
\usepackage{amssymb}
\usepackage{bm}
\usepackage{braket}
\usepackage{graphicx}
\usepackage{subcaption}

\newcommand{\unit}[1]{\,\mathrm{#1}}
\newcommand{\nl}{\nonumber\\&\qquad}

\begin{document}

\title{Study on axial fields in the dynamically assisted Schwinger effect}
\date{\today}
\author{Chengpeng Yu}
\email{yu.chengpeng@nucl.ph.tsukuba.ac.jp}
\affiliation{Center for Computational Sciences, University of Tsukuba, Tsukuba 305-8577, Japan}
\begin{abstract}
    The dynamically assisted Schwinger effect, the generation of fermion-antifermion pairs in vacuum under a strong, slow-varying field and a weak, high-frequency field, has become a promising avenue to probe the vacuum structure and the nonlinear dynamics of QED.
    However, the role of axial fields in this phenomenon has remained underexplored.
    This study aims at analyzing how spatial axial fields influence particle production in the dynamically assisted Schwinger effect.
    Employing the high-frequency effective theory based on the Floquet-Magnus expansion, we demonstrate that a spatial axial field can occur as the effective field of a circular polarized high-frequency plane wave and significantly increase the number of fermions produced across different timescales.
    This enhancement offers both theoretical insights and useful tools for the experimental implementation of the Schwinger effect.
\end{abstract}
\maketitle

\section{Introduction}
When an electric field is applied to vacuum, the vacuum generates pairs of fermions and antifermions \cite{Schwinger:1951nm}.
This effect, the Schwinger effect, is unique in QED.
Unlike the usual particle production in colliders, it does not require any on-shell particle in the initial state, neither photon nor charged particle.
Furthermore, the particle production rate of the Schwinger effect with a constant electric field satisfies $\Gamma\propto\exp\left(-\pi m^2/(e E)\right)$ \cite{Gelis:2015kya}, with $m$ as the electron mass, $E$ as the field strength, and $e$ as the electric charge.
(We use natural units $\hbar=c=1$.)
This behavior of $\Gamma$ is unexplainable from the naive framework of perturbative QED.
Therefore, the Schwinger effect reflects the instability of electroweak vacua and the nonlinearity of QED.
It receives continuous attention from the community, see \cite{Torgrimsson:2018xdf, Taya:2020dco, Yu:2023cic, Aleksandrov:2024rsz} for recent works.

Although important, experimental observation of the Schwinger effect is challenging.
According to $\Gamma\propto\exp\left(-\pi m^2/(e E)\right)$, the field strength threshold for the phenomenon to occur is $eE_{S} \sim m^2$.
In international system of units, this is about $10^{16}\unit{V/cm}$, far beyond the reach of current facilities \cite{Yoon:2021ony, Kohlfurst:2022edl, Hu:2023pmz}.
To lower the threshold, the community has, since 2008, explored the dynamically assisted Schwinger effect \cite{Schutzhold:2008pz, Dunne:2009gi, Torgrimsson:2016ant, Torgrimsson:2017pzs}.
The approach involves applying a weak, high-frequency electromagnetic field to the vacuum alongside the original electric field.
In this framework, the original field narrows the band gap between the Dirac sea and the positive-energy electron continuum in the vacuum; simultaneously, the high-frequency field excites electrons across the band gap, a process significantly more efficient than tunneling.
(See Fig. 19 in \cite{Gelis:2015kya} for the excitation mechanism and \cite{Wollert:2014epy} for the tunneling mechanism.)
Thus, even when the strength of the original field is below $m^2$, and the frequency of the high-frequency field is less than $m$, the particle production rate can still be substantial.
Based on the dynamically assisted Schwinger effect, one can drastically reduce the field strength threshold and implement the Schwinger effect with laser technology of the near future \cite{Taya:2020bcd, Ababekri:2019dkl, Fedotov:2013uja}.

The central idea of the dynamically assisted Schwinger effect, modulating a system using a high-frequency field, finds widespread application across various areas of physics.
In condensed matter physics, laser irradiation at frequencies much higher than the natural frequencies of a material is the common technique to modify the behavior of the material.
It is used to implement Floquet band gaps and anomalous Hall effect in graphene and related systems \cite{PhysRevResearch.4.013057, PhysRevB.79.081406, PhysRevB.99.214302, Luo:2023abp, Shi:2024rem}, to modulate the cold atom systems \cite{Mumford:2024yem, Wang:2024bbb}, and to eliminate the skin effect \cite{Chakrabarty:2024hpw} or induce Bose condensation \cite{Schnell:2022nhn} in non-Hermitian systems.
Beyond condensed matter physics, the configuration of low- and high- frequency fields also appears in domains including nuclear magnetic resonance, laser-plasma interaction, and geometric control of mechanical systems, see \cite{Blanes:2008xlr, Managa2016} for review.
Recently, this type of configuration has drawn attention in the high-energy community. 
For example, the authors of \cite{Yamada:2021jhy} studied the laser-driven chiral soliton lattice in the vacuum of QCD, and \cite{Fukushima:2023obj} studied the chiral magnetic effect under the high-frequency electric field.

Regardless of the specific physical context, an elegant approach to analyzing this type of configuration is the high-frequency effective theory.
This approach posits that when the frequency of a field is much higher than other natural frequencies of the system, one can view the high-frequency field as virtual processes dressing the low-frequency part of the Hamiltonian \cite{Bukov04032015}.
This leads to an effective Hamiltonian that is static on the timescale of the high-frequency field. 
The most well-established method for computing the effective Hamiltonian is the Floquet-Magnus expansion.
This method expresses the effective Hamiltonian as a series expansion in powers of $\omega^{-1}$, where $\omega$ is the frequency of the high-frequency field.
It is extensively discussed in the context of Floquet engineering, see \cite{Magnus:1954zz, PhysRevA.68.013820, annurev:/content/journals/10.1146/annurev-conmatphys-031218-013423, Goldman:2014xja, Aidelsburger:2017qlh} for details of this method. Alternative methods to derive the effective Hamiltonian are discussed in \cite{Eckardt:2015mtt, PhysRevLett.122.130604, RODRIGUEZVEGA2021168434}. Presumably, the concept of high-frequency effective theory is also applicable to the dynamically assisted Schwinger effect.

For the dynamically assisted Schwinger effect, the most intriguing aspect of high-frequency effective theories is their potential to induce artificial axial electromagnetic fields in the effective Hamiltonian.
Contrary to the ordinary electromagnetic fields, the axial fields act on left-handed and right-handed fermions with opposite signs, where the handedness is defined by the spin projection on the momentum direction. 
For effective axial fields in the Dirac and Weyl semimetals, see \cite{Ilan:2019lqk, Jamotte:2021mvq, Chernodub:2021nff, Cortijo:2015hlt, Arjona:2017qjc}; for axial fields in the relativistic matter with vortical fields or fluid helicity, see \cite{Gorbar:2016ygi, Landsteiner:2017hye, Yamamoto:2021gts}.
Recently, Copinger \textit{et al.} showed that a background axial field could drastically enhance the imaginary part of the Euler-Heisenberg Lagrangian of finite-mass fermions \cite{Copinger:2020nyx}.
This suggests that in the dynamically assisted Schwinger effect, if the high-frequency field induces an effective axial field, it can significantly magnify the particle production rate, especially in the regime where the field strength is weaker or close to the Schwinger threshold $E_S$, which is of particular interest from the experimental perspective.
Furthermore, in recent years, how helical fields influence the angular momentum of the fermions has become a hot topic in the study of the dynamically assisted Schwinger effect (see \cite{Huang:2019uhf, Aleksandrov:2024cqh, Kohlfurst:2022edl, Hu:2023pmz} for examples).
Since axial fields can be induced by high-frequency helical fields \cite{Yamada:2021jhy}, the picture of effective axial fields can provide a unique perspective for this research topic.
Therefore, the role of the axial field in a dynamically assist Schwinger effect is a topic both interesting for the experimentalists and theorists.

Despite its crucial role, the influence of axial fields on the Schwinger effect is currently a new topic.
For the temporal component of axial field $A_5^0$, one can identify it to the chiral chemical potential $\mu_5$ (see \cite{Fukushima:2008xe} for definition), and there are a few works on this aspect, see \cite{Domcke:2021fee}.
However, when it comes to the spatial components, the only paper we have observed so far is \cite{Copinger:2022bwl}, which focused on the general derivation of the QED one-loop effective action under a constant $\bm{A}_5$ as well as an electric (or magnetic) field.
To date, there has been no discussion on the high-frequency effective theory of the dynamically assisted Schwinger effect, which could give rise to axial fields, nor a systematic study on the influence of the axial fields on the particle production rate.

To fill this gap, in this paper, we plan to study the dynamically assisted Schwinger effect with an arbitrary static field and a high-frequency field that induces a spatial axial field.
Based on an effective Hamiltonian derived from the Floquet-Magnus expansion, we plan to derive a concise expression of the number of fermions produced in each state.
From this expression, we plan to figure out the relation among the high-frequency field, the effective axial field, and the fermions produced.
Finally, we plan to perform numerical calculations to show that the effective axial field enhances the particle production across different timescales.

The remaining part of the paper is organized as follows:
Sec.~\ref{sec:basics} reviews the canonical formulation of the Schwinger effect and the Floquet-Magnus expansion;
Sec.~\ref{sec:eff-theory} establishes the high-frequency effective theory of the dynamically assisted Schwinger effect;
Sec.~\ref{sec:long-time} derives the number of fermions in each state in the long-time limit;
Sec.~\ref{sec:axial} discusses how the axial field arises from the high-frequency field;
Sec.~\ref{sec:setup} discusses how to perform numerical calculation;
Sec.~\ref{sec:results} presents the results from numerical calculation and makes discussions;
and Sec.~\ref{sec:summary} summarizes this work.

\section{Basics of the Schwinger effect and the Floquet-Magnus expansion}
\label{sec:basics}

In this section, we review two aspects of background knowledge essential for constructing our theoretical model.
The first is the canonical formulation of the Schwinger effect to compute the number of fermions produced in each state.
The second is the high-frequency effective theory based on Floquet-Magnus expansion.

\subsection{Schwinger effect}

We consider the Schwinger effect under time-dependent external fields. 
The single-particle Hamiltonian under the external fields is $\hat{H}(t)$.
We assume the external fields occur at $t>t_{\text{in}}$.
Before that, $\hat{H}(t)$ equals the free-particle Dirac Hamiltonian.

At $t=t_{\text{in}}$, the positive- and negative-energy eigenmodes of $\hat{H}(t)$, $\ket{u_{\text{in}}^{\alpha}(t_{\text{in}})}$ and $\ket{v_{\text{in}}^{\alpha}(t_{\text{in}})}$, satisfies
\begin{equation}
    \hat{H}(t_{\text{in}}) \ket{u_{\text{in}}^{\alpha}(t_{\text{in}})} = \epsilon_{\text{in}}^{\alpha} \ket{u_{\text{in}}^{\alpha}(t_{\text{in}})},
\end{equation}
\begin{equation}
    \hat{H}(t_{\text{in}}) \ket{v_{\text{in}}^{\alpha}(t_{\text{in}})} = -\bar{\epsilon}_{\text{in}}^{\alpha} \ket{v_{\text{in}}^{\alpha}(t_{\text{in}})},
\end{equation}
where $\epsilon_{\text{in}}^{\alpha}>0$, $\bar{\epsilon}_{\text{in}}^{\alpha}>0$.
$\alpha$ is the index of different eigenmodes.
Here, we require all the eigenmodes to be orthonormal to each other, i.e.,
\begin{equation}
    \braket{u_{\text{in}}^{\alpha}(t_{\text{in}})|u_{\text{in}}^{\beta}( t_{\text{in}})}
    =\delta^{\alpha\beta},
\end{equation}
\begin{equation}
    \braket{v_{\text{in}}^{\alpha}(t_{\text{in}})|v_{\text{in}}^{\beta}(t_{\text{in}})}=\delta^{\alpha\beta},
\end{equation}
\begin{equation}
    \braket{u_{\text{in}}^{\alpha}(t_{\text{in}})|v_{\text{in}}^{\beta}(t_{\text{in}})}=0,
\end{equation}
where $\braket{u_{\text{in}}^{\alpha}(t_{\text{in}})|u_{\text{in}}^{\beta}( t_{\text{in}})} =\int d^{3}\bm{x}\,u_{\text{in}}^{\alpha\dagger}(\bm{x},t_{\text{in}})u_{\text{in}}^{\beta}(\bm{x},t_{\text{in}})$, $\braket{\bm{x}|u_{\text{in}}^{\beta}( t_{\text{in}})}=u_{\text{in}}^{\beta}(\bm{x},t_{\text{in}})$.

These eigenmodes evolve under time-evolution operator $\hat{U}(t,t_{\text{in}})=\mathcal{T}\exp(-i\int_{t_{\text{in}}}^{t}d\tau\,\hat{H}(\tau))$.
We define
\begin{equation}
    \ket{u_{\text{in}}^{\alpha}(t)}=\hat{U}(t,t_{\text{in}})\ket{u_{\text{in}}^{\alpha}(t_\text{in})},
\end{equation}
\begin{equation}
    \ket{v_{\text{in}}^{\alpha}(t)}=\hat{U}(t,t_{\text{in}})\ket{u_{\text{in}}^{\alpha}(t_\text{in})}.
\end{equation}
Here, $\ket{u_{\text{in}}^{\alpha}(t)}$ and $\ket{v_{\text{in}}^{\alpha}(t)}$ are not necessarily eigenmodes of $\hat{H}(t)$.
However, since $\hat{U}(t,t_{\text{in}})$ is Hermitian, $\ket{u_{\text{in}}^{\alpha}(t)}$ and $\ket{v_{\text{in}}^{\alpha}(t)}$ follow the same orthonormal relations as $\ket{u_{\text{in}}^{\alpha}(t_{\text{in}})}$ and $\ket{v_{\text{in}}^{\alpha}(t_{\text{in}})}$, i.e., $\braket{u_{\text{in}}^{\alpha}(t)|u_{\text{in}}^{\beta}(t)}=\delta^{\alpha\beta}$, $\braket{v_{\text{in}}^{\alpha}(t)|v_{\text{in}}^{\beta}(t)}=\delta^{\alpha\beta}$, and $\braket{u_{\text{in}}^{\alpha}(t)|v_{\text{in}}^{\beta}(t)}=0$.

With these eigenmodes, we construct the Dirac field operator under the external fields.
This type of field operators is often referred to as the field operators in the Furry picture, see \cite{Furry:1951bef, Moortgat-Pick:2009fyg} for introduction.
We can either express it in a Schr\"odinger way,
\begin{equation}
    \psi(\bm{x},t) =\sum_{\alpha}\left(b_{\text{in}}^{\alpha}u_{\text{in}}^{\alpha}(\bm{x},t)+c_{\text{in}}^{\alpha\dagger}v_{\text{in}}^{\alpha}(\bm{x},t)\right),
    \label{eq:psi-Schro}
\end{equation}
or Heisenberg way,
\begin{equation}
    \psi(\bm{x},t)=\sum_{\alpha}\left(b^{\alpha}_{\text{in}}(t)u_{\text{in}}^{\alpha}(\bm{x},t_{\text{in}})+c^{\alpha\dagger}_{\text{in}}(t)v_{\text{in}}^{\alpha}(\bm{x},t_{\text{in}})\right).
    \label{eq:psi:Heisen}
\end{equation}
Here, $b_{\text{in}}^{\alpha}$ and $c_{\text{in}}^{\alpha}$ are the annihilation operators of the fermions and antifermions at $t=t_{\text{in}}$, respectively.
Since the external fields vanish at $t \le t_{\text{in}}$, they are identical to the annihilation operators of free fermions.
$b^{\alpha}_{\text{in}}(t)$ and $c^{\alpha}_{\text{in}}(t)$ are the time-dependent annihilation operators satisfying $b^{\alpha}_{\text{in}}(t_{\text{in}}) = b_{\text{in}}^{\alpha}$, $c^{\alpha}_{\text{in}}(t_{\text{in}}) = c_{\text{in}}^{\alpha}$.
Equating Eq.~\eqref{eq:psi-Schro} with Eq.~\eqref{eq:psi:Heisen} and using the orthonormal relations, we prove the following Bogoliubov transformation that connects $b^{\alpha}_{\text{in}}(t)$, $c^{\alpha\dagger}_{\text{in}}(t)$ and $b_{\text{in}}^{\alpha}$, $c_{\text{in}}^{\alpha\dagger}$:
\begin{equation}
    b^{\alpha}_{\text{in}}(t)=\sum_{\beta}\left(b_{\text{in}}^{\beta}\braket{u_{\text{in}}^{\alpha}(t_{\text{in}})|u_{\text{in}}^{\beta}(t)}+c_{\text{in}}^{\beta\dagger}\braket{u_{\text{in}}^{\alpha}(t_{\text{in}})|v_{\text{in}}^{\beta}(t)}\right),
    \label{eq:Bogo-b}
\end{equation}
\begin{equation}
    c^{\alpha\dagger}_{\text{in}}(t)=\sum_{\alpha}\left(b_{\text{in}}^{\beta}\braket{v_{\text{in}}^{\alpha}(t_{\text{in}})|u_{\text{in}}^{\beta}(t)}+c_{\text{in}}^{\beta\dagger}\braket{v_{\text{in}}^{\alpha}(t_{\text{in}})|v_{\text{in}}^{\beta}(t)}\right).
\end{equation}

At arbitrary time $t$, the average number of positive-energy fermions in eigenstate $\alpha$ is
\begin{equation}
    n^{\alpha}(t) = \braket{0_{\text{in}}|b^{\alpha\dagger}_{\text{in}}(t)b^{\alpha}_{\text{in}}(t)|0_{\text{in}}}.
\end{equation}
Here, $\ket{0_{\text{in}}}$ is the vacuum state at $t=t_\text{in}$, defined by $b_{\text{in}}^{\alpha}\ket{0_{\text{in}}}=0$, $c_{\text{in}}^{\alpha}\ket{0_{\text{in}}}=0$.
Substituting Eq.~\eqref{eq:Bogo-b} into the above expression, we have
\begin{equation}
    n^{\alpha}(t) = \sum_{\beta}\left|\braket{u_{\text{in}}^{\alpha}(t_{\text{in}})|v_{\text{in}}^{\beta}(t)}\right|^{2}.
    \label{eq:n-genreal}
\end{equation}
Hence, once we solve the Dirac equation with the external fields and obtain $\ket{u^{\alpha}_{\text{in}}(t)}$, $\ket{v^{\beta}_{\text{in}}(t)}$, we can compute the time-dependent fermion number produced by the Schwinger effect. 

\subsection{Floquet-Magnus expansion}

Let us consider the solution of a general single-particle equation of motion,
\begin{equation}
    i\frac{\partial}{\partial t} \psi(\bm{x}, t) = \hat{H}(t) \psi(\bm{x},t),
\end{equation}
where $\hat{H}(t)$ is an arbitrary Hamiltonian with time periodicity $\hat{H}(t)=\hat{H}(t+2\pi/\omega)$.
The time-evolution operator $\hat{U}(t,t_{\text{in}})$ of $\psi(\bm{x}, t)$ satisfies
\begin{equation}
    i\frac{\partial}{\partial t} \hat{U}(t,t_{\text{in}}) = \hat{H}(t) \hat{U}(t,t_{\text{in}}).
    \label{eq:EOM-U}
\end{equation}

The Floquet-Magnus expansion uses the following ansatz of $\hat{U}(t,t_\text{in})$:
\begin{equation}
    \hat{U}(t,t_{\text{in}}) = e^{-i\hat{K}(t)} e^{-i \hat{H}_F (t-t_{\text{in}})} e^{i\hat{K}(t_{\text{in}})},
    \label{eq:FM-ansatz}
\end{equation}
where $\hat{H}_F$ is time independent, $\hat{K}(t)$ satisfies $\hat{K}(t)=\hat{K}(t+2\pi/\omega)$.
Both operators are Hermitian.
When $\omega$ is high, $\hat{K}(t)$ represents the small-magnitude, rapid oscillation (micromotion) of the system, and $\hat{H}_F$ represents the average motion.
Hence, they are referred to as the kick operator and the Floquet effective Hamiltonian, respectively.

From Eq.~\eqref{eq:FM-ansatz}, $\hat{K}(t)$ is not uniquely fixed.
For example, if $\hat{K}(t)$ satisfies Eq.~\eqref{eq:FM-ansatz}, $\hat{K}'(t)=\hat{K}(t)+C$ satisfies it as well (where $C$ is an arbitrary commuting number).
For the convenience of latter derivation, from then on we require
\begin{equation}
    \int dt \hat{K}(t) = 0.
\end{equation}
In literature, the Floquet-Magnus expansion with this convention is often referred to as the van Vleck expansion, for example, see \cite{Managa2016, RODRIGUEZVEGA2021168434, Eckardt:2015mtt}.

The Floquet-Magnus expansion expands $\hat{H}_F$ and $\hat{K}(t)$ as series in powers of $\omega$,
\begin{equation}
    \hat{H}_F = \sum_{n=0}^{+\infty} \hat{H}_F^{(n)},
    \label{eq:H-power}
\end{equation}
\begin{equation}
    \hat{K}(t) = \sum_{n=0}^{+\infty} \hat{K}^{(n)}(t),
    \label{eq:K-power}
\end{equation}
where $\hat{H}_F^{(n)}\propto\omega^{-n}$, $\hat{K}^{(n)}(t)\propto\omega^{-n}$, $\hat{K}^{(n)}(t)=\hat{K}^{(n)}(t+2\pi/\omega)$, and $\hat{K}^{(n)}(0)=0$.
Substitute Eq.~\eqref{eq:FM-ansatz} and the above power series into Eq.~\eqref{eq:EOM-U}. After a series of algebras, one obtains the  order-by-order expressions of $\hat{H}_F^{(n)}$ and $\hat{K}^{(n)}(t)$. Up to the second order, the results are
\begin{equation}
    \hat{H}_F^{(0)} = \tilde{H}_0,
    \label{eq:HF-0-gen}
\end{equation}
\begin{equation}
    \hat{H}_F^{(1)} = \frac{1}{2} \sum_{l\neq 0} \frac{1}{l\omega} \left[
        \tilde{H}_l, \tilde{H}_{-l}
    \right],
    \label{eq:HF-1}
\end{equation}
\begin{align}
    &\hat{H}_F^{(2)} = \frac{1}{2} \sum_{l\neq 0} \left( \frac{1}{il\omega}\right)^2
    \left[
        \tilde{H}_{-l}, \left[ \tilde{H}_l, \tilde{H}_0\right]
    \right]\nl
    - \frac{1}{3} \sum_{l\neq 0} \sum_{l'\neq 0,-l}
    \frac{1}{il'\omega} \frac{1}{il\omega}
    \left[
        \tilde{H}_{l}, \left[ \tilde{H}_{l'}, \tilde{H}_{-l-l'}\right]
    \right],
\end{align}
\begin{equation}
    \hat{K}^{(0)}(t) = 0,
    \label{eq:K-0}
\end{equation}
\begin{equation}
    \hat{K}^{(1)}(t) = \sum_{l\neq 0} \frac{1}{il\omega} \tilde{H}_l e^{il\omega t},
    \label{eq:K-1}
\end{equation}
\begin{align}
    &\hat{K}^{(2)}(t) = \frac{i}{2} \sum_{l\neq 0} \sum_{l'\neq 0,l}
    \frac{1}{il'\omega} \frac{1}{il\omega} e^{il\omega t}
    \left[\tilde{H}_{l'}, \tilde{H}_{l-l'}\right]\nl
    + i \sum_{l\neq 0} \left( \frac{1}{il\omega}\right)^2 e^{il\omega t}
    \left[ \tilde{H}_l, \tilde{H}_0\right].
\end{align}
Here, $\tilde{H}_l$ is
\begin{equation}
    \tilde{H}_{l} = \frac{2\pi}{\omega} \int_{2\pi/\omega} dt \, \hat{H}(t) e^{il\omega t}.
\end{equation}
For a detailed derivation of the above expressions, we refer the readers to \cite{Fukushima:2023obj}.

\section{The high-frequency effective theory}
\label{sec:eff-theory}

Let us apply the Floquet-Magnus expansion to study the Schwinger effect.
As in the previous section, we denote the single-particle Hamiltonian under the external field as $\hat{H}(t)$, and denote the corresponding Floquet effective Hamiltonian and kick operator as $\hat{H}_F$ and $\hat{K}(t)$, respectively.

We denote the positive-energy and negative-energy eigenmodes of $\hat{H}_F$ as $\ket{u_{F}^{\alpha}}$ and $\ket{v_{F}^{\alpha}}$, respectively.
They satisfy
\begin{equation}
    \hat{H}_{F} \ket{u_{F}^{\alpha}} = \epsilon_{F}^{\alpha} \ket{u_{F}^{\alpha}},
    \label{eq:u-EoM-F}
\end{equation}
\begin{equation}
    \hat{H}_{F} \ket{v_{F}^{\alpha}} = -\bar{\epsilon}_{F}^{\alpha} \ket{v_{F}^{\alpha}},
    \label{eq:v-EoM-F}
\end{equation}
where $\epsilon_F^{\alpha}>0$, $\bar{\epsilon}_F^{\alpha}>0$.
$\alpha$ is the index of different eigenmodes.
Furthermore, we require the eigenmodes to satisfy the orthonormal relations $\braket{u_F^{\alpha}|u_F^{\beta}}=\delta^{\alpha\beta}$, $\braket{v_F^{\alpha}|v_F^{\beta}}=\delta^{\alpha\beta}$, and $\braket{u_F^{\alpha}|v_F^{\beta}}=0$, where $\braket{u_F^{\alpha}|u_F^{\beta}} =\int d^{3}\bm{x}\,u_F^{\alpha\dagger}(\bm{x}) u_F^{\beta}(\bm{x})$ and $\braket{\bm{x}|u_F^{\beta}}=u_F^{\beta}(\bm{x})$.
Because one can choose a different basis of the eigenspaces, $\ket{u_{F}^{\alpha}}$ and $\ket{v_{F}^{\alpha}}$ are not uniquely fixed by the above requirements.
The seemingly natural expression $\ket{u_{\text{in}}^{\alpha}(t_{\text{in}})} = e^{-i\hat{K}(t_{\text{in}})} \ket{u_{F}^{\alpha}}$, $\ket{v_{\text{in}}^{\alpha}(t_{\text{in}})} = e^{-i\hat{K}(t_{\text{in}})} \ket{v_{F}^{\alpha}}$ does not necessarily hold.

The Floquet-Magnus expansion guarantees that $\ket{u_{F}^{\alpha}}$ and $\ket{v_{F}^{\alpha}}$ satisfy the completeness relation:
\begin{equation}
    \sum_{\alpha}\left(\ket{u_{F}^{\alpha}} \bra{u_{F}^{\alpha}} 
    +\ket{v_{F}^{\alpha}} \bra{v_{F}^{\alpha}} \right) = I,
\end{equation}
with $I$ as the identity operator.
Plugging this relation into $\ket{v_{\text{in}}^{\alpha}(t)} = \hat{U}(t,t_{\text{in}}) \ket{u_{\text{in}}^{\alpha}(t_\text{in})}$, we have
\begin{align}
    &\ket{v_{\text{in}}^{\beta}(t)} = e^{-i\hat{K}(t)} e^{-i\hat{H}_{F}(t-t_{\text{in}})} e^{i\hat{K}(t_{\text{in}})} \ket{v_{\text{in}}^{\beta}(t_{\text{in}})}\nl
    = e^{-i\hat{K}(t)} e^{-i\hat{H}_{F}(t-t_{\text{in}})} \nl
    \qquad \sum_{\gamma} \biggl[ \braket{u_{F}^{\gamma}| e^{i\hat{K}(t_{\text{in}})}|v_{\text{in}}^{\beta}(t_{\text{in}})} \ket{u_{F}^{\gamma}}\nl
    \qquad +\braket{v_{F}^{\gamma}|e^{i\hat{K}(t_{\text{in}})}|v_{\text{in}}^{\beta}(t_{\text{in}})} \ket{v_{F}^{\gamma}}\biggr]\nl
    =e^{-i\hat{K}(t)} \nl
    \qquad \sum_{\gamma} \biggl[ e^{-i\epsilon_{F}^{\gamma}(t-t_{\text{in}})} \braket{u_{F}^{\gamma}| e^{i\hat{K}(t_{\text{in}})}|v_{\text{in}}^{\beta}(t_{\text{in}})} \ket{u_{F}^{\gamma}} \nl
    \qquad+ e^{i\bar{\epsilon}_{F}^{\gamma}(t-t_{\text{in}})} \braket{v_{F}^{\gamma}|e^{i\hat{K}(t_{\text{in}})}|v_{\text{in}}^{\beta}(t_{\text{in}})} \ket{v_{F}^{\gamma}} \biggr].
\end{align}

With the expression of $\ket{v_{\text{in}}^{\beta}(t)}$, we rewrite Eq.~\eqref{eq:n-genreal} as
\begin{align}
    n^{\alpha}(t)=\sum_{\beta}\left|\braket{u_{\text{in}}^{\alpha}(t_{\text{in}})|e^{-i\hat{K}(t)}\hat{S}(t)e^{i\hat{K}(t_{\text{in}})}|v_{\text{in}}^{\beta}(t_{\text{in}})}\right|^{2},
    \label{eq:n-t}
\end{align}
where
\begin{equation}
    \hat{S}(t)=\sum_{\gamma}\left(\ket{u_{F}^{\gamma}} \bra{u_{F}^{\gamma}} e^{-i\epsilon_{F}^{\gamma}(t-t_{\text{in}})}
    + \ket{v_{F}^{\gamma}} \bra{v_{F}^{\gamma}} e^{i\bar{\epsilon}_{F}^{\gamma}(t-t_{\text{in}})}\right).
    \label{eq:S}
\end{equation}

The above expression gives the number of positive-energy fermions in each state from the high-frequency effective theory.
It is a critical result because, in this expression, the fermion number depends only on $\ket{u_{\text{in}}^{\alpha}(t_{\text{in}})}$, $\ket{v_{\text{in}}^{\alpha}(t_{\text{in}})}$, $\ket{u_{F}^{\alpha}}$, $\ket{v_{F}^{\alpha}}$ and the corresponding energies.
In our setup, $\ket{u_{\text{in}}^{\alpha}(t_{\text{in}})}$ and $\ket{v_{\text{in}}^{\alpha}(t_{\text{in}})}$ are nothing but the eigenmodes of the free-particle Dirac Hamiltonian.
$\ket{u_{F}^{\alpha}}$ and $\ket{v_{F}^{\alpha}}$ are the eigenmodes of $\hat{H}_F$ that is static.
Therefore, we transform the problem of solving the Schwinger effect under time-dependent fields to a static problem.
This significantly simplifies the analysis of the dynamically assisted Schwinger effect.

To verify the validity of the above expression, we consider two extreme cases.
The first case is the $n^{\alpha}(t)$ at $t=t_{\text{in}}$.
At this situation, one can show $\hat{S}(t)=I$ because of the completeness relation, so the $e^{-i\hat{K}(t)}$ factor cancels with the $e^{i\hat{K}(t_{\text{in}})}$ factor and
\begin{equation}
    n^{\alpha}(t_{\text{in}}) = \sum_{\beta}\left|\braket{u_{\text{in}}^{\alpha}(t_{\text{in}})|v_{\text{in}}^{\beta}(t_{\text{in}})}\right|^2
    =0.
\end{equation}
This means that when an external field is applied, the fermions do not occur all of a sudden; instead, the fermion density increases smoothly, which is the expected behavior.
The second case is of vanishing external field.
At this case, we have $\hat{K}(t)=0$ and $\hat{H}(t_{\text{in}})=\hat{H}_F$.
Since $\hat{H}(t_{\text{in}})$ and $\hat{H}_F$ share the same eigenspaces, without loss of generality, we can choose the bases such that $\ket{u_{F}^{\alpha}}=\ket{u_{\text{in}}^{\alpha}(t_{\text{in}})}$, $\ket{v_{F}^{\alpha}}=\ket{v_{\text{in}}^{\alpha}(t_{\text{in}})}$ with $\epsilon_{\text{in}}^{\alpha}=\epsilon_{F}^{\alpha}$ and $\bar{\epsilon}_{\text{in}}^{\alpha}=\bar{\epsilon}_{F}^{\alpha}$.
This yields
\begin{align}
    &n^{\alpha}(t)=\biggl|\sum_{\gamma}\biggl(\braket{u_{\text{in}}^{\alpha}(t_{\text{in}})|u_{\text{in}}^{\gamma}(t_{\text{in}})} \braket{u_{\text{in}}^{\gamma}(t_{\text{in}})|v_{\text{in}}^{\beta}(t_{\text{in}})}e^{-i\epsilon_{\text{in}}^{\gamma}(t-t_{\text{in}})}\nl
    \qquad+\braket{u_{\text{in}}^{\alpha}(t_{\text{in}})|v_{\text{in}}^{\gamma}(t_{\text{in}})} \braket{v_{\text{in}}^{\gamma}(t_{\text{in}})|v_{\text{in}}^{\beta}(t_{\text{in}})} e^{i\bar{\epsilon}_{\text{in}}^{\gamma}(t-t_{\text{in}})}\biggr)\biggr|^{2}\nl
    =0,
\end{align}
which is also the expected results.

\section{The long-time limit}
\label{sec:long-time}

From the experimental perspective, the average value of $n^{\alpha}(t)$ at large $t$ is an important observable.
In this section, we derive an expression of it.

We rewrite $\hat{S}(t)$ as
\begin{equation}
    \hat{S}(t)=\sum_{\epsilon_{F}} \hat{S}^{(+)}(\epsilon_{F}) e^{-i\epsilon_{F}(t-t_{\text{in}})}
    +\sum_{\bar{\epsilon}_{F}} \hat{S}^{(-)}(\bar{\epsilon}_{F})e^{i\bar{\epsilon}_{F}(t-t_{\text{in}})},
\end{equation}
\begin{equation}
    \hat{S}^{(+)}(\epsilon_{F})=\sum_{\gamma,\epsilon_{F}^{\gamma}=\epsilon_{F}} \ket{u_{F}^{\gamma}} \bra{u_{F}^{\gamma}},
    \label{eq:Splus}
\end{equation}
\begin{equation}
    \hat{S}^{(-)}(\bar{\epsilon}_{F})= \sum_{\gamma,\bar{\epsilon}_{F}^{\gamma}=\bar{\epsilon}_{F}} \ket{v_{F}^{\gamma}} \bra{v_{F}^{\gamma}}.
    \label{eq:Sminus}
\end{equation}
Here, the summations $\sum_{\gamma,\epsilon_{F}^{\gamma}=\epsilon_{F}}$, $\sum_{\gamma,\bar{\epsilon}_{F}^{\gamma}=\bar{\epsilon}_{F}}$ sum up degenerate eigenstates of $\hat{H}_F$ with energy $\epsilon_F$ or $\bar{\epsilon}_F$, respectively.

When discussing the long-time behavior of $n^{\alpha}(t)$, we are not interested in the micromotion generated by the kick operator. Hence, we choose $\hat{K}(t)\simeq 0$, $\hat{K}(t_\text{in})\simeq 0$, and expand the expression of $n^{\alpha}(t)$ with $\hat{S}^{(+)}(\epsilon_{F})$ and $\hat{S}^{(-)}(\bar{\epsilon}_{F})$ as
\begin{align}
    &n^{\alpha}(t)=\sum_{\beta}\biggl(\sum_{\epsilon_{F}}\sum_{\epsilon_{F}'}
        \braket{u_{\text{in}}^{\alpha}(t_{\text{in}})|\hat{S}^{(+)}(\epsilon_{F})|v_{\text{in}}^{\beta}(t_{\text{in}})}\nl
        \qquad\times\braket{v_{\text{in}}^{\beta}(t_{\text{in}})|\hat{S}^{(+)\dagger}(\epsilon_{F}')|u_{\text{in}}^{\alpha}(t_{\text{in}})}
        e^{-i(\epsilon_{F}-\epsilon_{F}')(t-t_{\text{in}})}\nl
    +\sum_{\epsilon_{F}}\sum_{\bar{\epsilon}_{F}'}
        \braket{u_{\text{in}}^{\alpha}(t_{\text{in}})|\hat{S}^{(+)}(\epsilon_{F})|v_{\text{in}}^{\beta}(t_{\text{in}})}\nl
        \qquad\times\braket{v_{\text{in}}^{\beta}(t_{\text{in}})|\hat{S}^{(-)\dagger}(\bar{\epsilon}_{F}')|u_{\text{in}}^{\alpha}(t_{\text{in}})}
        e^{-i(\epsilon_{F}+\bar{\epsilon}_{F}')(t-t_{\text{in}})}\nl
    +\sum_{\bar{\epsilon}_{F}}\sum_{\epsilon_{F}'}
        \braket{u_{\text{in}}^{\alpha}(t_{\text{in}})|\hat{S}^{(-)}(\bar{\epsilon}_{F})|v_{\text{in}}^{\beta}(t_{\text{in}})}\nl
        \qquad\times\braket{v_{\text{in}}^{\beta}(t_{\text{in}})|\hat{S}^{(+)\dagger}(\epsilon_{F}')|u_{\text{in}}^{\alpha}(t_{\text{in}})}
        e^{i(\bar{\epsilon}_{F}+\epsilon_{F}')(t-t_{\text{in}})}\nl
    +\sum_{\bar{\epsilon}_{F}}\sum_{\bar{\epsilon}_{F}'}
        \braket{u_{\text{in}}^{\alpha}(t_{\text{in}})|\hat{S}^{(-)}(\bar{\epsilon}_{F})|v_{\text{in}}^{\beta}(t_{\text{in}})}\nl
        \qquad\times\braket{v_{\text{in}}^{\beta}(t_{\text{in}})|\hat{S}^{(-)\dagger}(\bar{\epsilon}_{F}')|u_{\text{in}}^{\alpha}(t_{\text{in}})}
        e^{i(\bar{\epsilon}_{F}-\bar{\epsilon}_{F}')(t-t_{\text{in}})}\biggr).
    \label{eq:n-no-kick}
\end{align}

To proceed further, let us consider the long-time behavior of $n^{\alpha}(t)$ under a constant electric field.
From \cite{Gelis:2015kya}, we know in that case, the $n^{\alpha}(t)$ stabilizes at $n^{\alpha} = \exp\left(-\pi (m^2+\bm{p}_T^2)/(e E)\right)$, where $E$ is the electric field strength, and $\bm{p}_T$ is the momentum in the transverse direction of the electric field.
This behavior occurs as the following two processes balance each other:
(1) increase of the fermion number from the particle production, and
(2) decrease of the fermion number when the old fermions are accelerated by the field to higher-energy states.
This balancing mechanism also occurs in our system described by $\hat{H}_F$ that is static, so stabilized $n^{\alpha}(t)$ occurs in the long time as well.
From the above consideration, the only relevant terms in Eq.~\eqref{eq:n-no-kick} are the first and last ones with $\epsilon_F=\epsilon_F'$, $\bar{\epsilon}_F=\bar{\epsilon}_F'$, respectively.
The number of positive-energy fermions in state $\alpha$ in the long-time limit becomes
\begin{align}
    &n^{\alpha}=\sum_{\beta}
    \biggl[\sum_{\epsilon_{F}} \braket{u_{\text{in}}^{\alpha}(t_{\text{in}})|\hat{S}^{(+)}(\epsilon_{F})|v_{\text{in}}^{\beta}(t_{\text{in}})}\nl
    \qquad
    \braket{v_{\text{in}}^{\beta}(t_{\text{in}})|\hat{S}^{(+)\dagger}(\epsilon_{F})|u_{\text{in}}^{\alpha}(t_{\text{in}})}\nl
    +\sum_{\bar{\epsilon}_{F}} \braket{u_{\text{in}}^{\alpha}(t_{\text{in}})|\hat{S}^{(-)}(\bar{\epsilon}_{F})|v_{\text{in}}^{\beta}(t_{\text{in}})}\nl
    \qquad
    \braket{v_{\text{in}}^{\beta}(t_{\text{in}})|\hat{S}^{(-)\dagger}(\bar{\epsilon}_{F})|u_{\text{in}}^{\alpha}(t_{\text{in}})}\biggr].
    \label{eq:n-long-time}
\end{align}

\section{Arise of the axial field}
\label{sec:axial}

Let us consider how axial fields arise from the high-frequency effective theory.
From then on, we consider an external field $A_{\mu}(x)$ that consists of a static component and a high-frequency component of frequency $\omega$.
We define the static component as
\begin{equation}
    \bar{A}_{\mu}(\bm{x})=\frac{\omega}{2\pi}\int_{2\pi/\omega}dt\,A_{\mu}(x) ,
\end{equation}
and the high-frequency component as
\begin{equation}
    \tilde{A}_{\mu}(\bm{x}) = \frac{\omega}{2\pi}\int_{2\pi/\omega}dt\,A_{\mu}(x) e^{i\omega t}.
\end{equation}
With the external field, the single-particle Dirac Hamiltonian becomes
\begin{align}
    &\hat{H}(t) = -i\gamma^{0}\bm{\gamma}\cdot\bm{\nabla}+m\gamma^{0}\nl
    + e\gamma^{0}\gamma^{\mu} \left(
        \bar{A}_{\mu}(\bm{x})
        + \tilde{A}_{\mu}(\bm{x})e^{-i\omega t}
        + \tilde{A}_{\mu}^\dagger(\bm{x})e^{i\omega t}
    \right).
\end{align}
Here, $\mu=0,1,2,3$ and we use the Einstein summation notion.
For 3D vectors like $\bm{\gamma}$, we use the convention $\bm{\gamma}=(\gamma^1, \gamma^2, \gamma^3)^T$.
For the gradient operator $\bm{\nabla}$, we use the convention $\bm{\nabla}=(\partial_x, \partial_y, \partial_z)^T$.

We conduct the first-order Floquet-Magnus expansion of the above Hamiltonian. From Eqs.~\eqref{eq:HF-0-gen}, \eqref{eq:HF-1}, \eqref{eq:K-0}, and \eqref{eq:K-1}, the result is
\begin{equation}
    \hat{H}_{F}^{(0)}=-i\gamma^{0}\bm{\gamma}\cdot\bm{\nabla}+m\gamma^{0}+e\gamma^{0}\gamma^{\mu}\bar{A}_{\mu}(\bm{x}),
    \label{eq:HF-0}
\end{equation}
\begin{equation}
    \hat{H}_{F}^{(1)}=-\frac{e^{2}}{\omega}[\gamma^{0}\gamma^{\mu},\gamma^{0}\gamma^{\nu}]\tilde{A}_{\mu}(\bm{x})\tilde{A}_{\nu}^{*}(\bm{x}),
    \label{eq:HF-1-interm}
\end{equation}
\begin{equation}
    \hat{K}^{(0)}(t)=0,
\end{equation}
\begin{equation}
    \hat{K}^{(1)}(t)=\frac{1}{-i\omega}e\gamma^{0}\gamma^{\mu}\left(\tilde{A}_{\mu}(\bm{x})e^{-i\omega t}-\tilde{A}_{\mu}^{*}(\bm{x})e^{i\omega t}\right).
\end{equation}

From the above expansion, one interesting observation is that, up to the first order, $\exp(-i\hat{K}(t))$ is
\begin{equation}
    e^{-i\hat{K}(t)}=e^{-ie\gamma^{0}\gamma^{\mu}\int dt\,\left(\tilde{A}_{\mu}(\bm{x})e^{-i\omega t}
    +\tilde{A}_{\mu}^*(\bm{x})e^{i\omega t}\right)}.
\end{equation}
This expression shows that if we treat the high-frequency part of the external field, $\tilde{A}_{\mu}(\bm{x})\exp(-i\omega t)+\tilde{A}_{\mu}^*(\bm{x})\exp(i\omega t)$, as a perturbation, then $\exp(-i\hat{K}(t))$ is identical to the single-particle time-evolution operator in the interaction picture.
Therefore, one can interpret the leading-order result of the high-frequency effective theory as an extension of the perturbative QED.

More importantly, after doing some Dirac algebra (see the Appendix), we transform Eq.~\eqref{eq:HF-1-interm} into
\begin{equation}
    \hat{H}_{F}^{(1)}=-2i\frac{e^{2}}{\omega}\gamma^{0}\bm{\gamma}\gamma^{5}\cdot\left(\tilde{\bm{A}}(\bm{x})\times\tilde{\bm{A}}^{*}(\bm{x})\right).
    \label{eq:HF-A5}
\end{equation}
Here, convention of the cross product is $(\tilde{\bm{A}}\times\tilde{\bm{A}}^{*})^{i}=\epsilon^{ijk}\tilde{A}_j\tilde{A}^*_k$, where $i,j,k\in\{1,2,3\}$ and $\epsilon^{ijk}$ is the Levi-Civita symbol with $\epsilon^{123}=1$.

In a Dirac Hamiltonian, the interaction term between fermions and an axial electromagnetic field $A_5^{\mu}(x)$ is $e\gamma^0\gamma^{\mu}\gamma^5 A_{5\mu}$.
Comparing this expression with Eq.~\eqref{eq:HF-A5}, we identify $H_{F}^{(1)}$ to the coupling term with the following effective axial field:
\begin{equation}
    A_{5}^0 = 0,
\end{equation}
\begin{equation}
    \bm{A}_{5}(\bm{x})=\frac{2ie}{\omega}\left(\tilde{\bm{A}}(\bm{x})\times\tilde{\bm{A}}^{*}(\bm{x})\right).
    \label{eq:A5}
\end{equation}
Thus, the high-frequency effective theory induces an effective spatial axial field in the system.
This field is interesting because under parity transformation, $\gamma^5 \to -\gamma^5$. 
Thus, the axial field and the vector field in the system have opposite parity, and the combination of them can violate the parity symmetry of the system.

To show the exact form of $\bm{A}_{5}(\bm{x})$, let us discuss a specific field distribution that is easy to implement in experiments.
We consider a static electric field and a circular polarized high-frequency plane wave.
Under coulomb gauge, we have
\begin{equation}
    A^{0}(x) = \varphi(\bm{x}),
    \label{eq:A0}
\end{equation}
\begin{equation}
    \bm{A}(x) =
    (\bm{e}_{1}-i\bm{e}_{2})f(\bm{x})e^{i\omega (z-t)}
    + \mathrm{c.c.}
    \label{eq:A-space}
\end{equation}
Here, $\varphi(\bm{x})$ is the scalar potential corresponding to the static electric field, $\bm{e}_1$, $\bm{e}_2$ are the unit vectors in $x$ and $y$ directions, and $f(\bm{x})$ is a slow-varying envelope of the plane wave.

Under this field, one can show $\bar{A}_{\mu}(\bm{x}) = e_{\mu}^0 \varphi(\bm{x})$ and $\tilde{\bm{A}}(\bm{x})=(\bm{e}_{1}-i\bm{e}_{2})f(\bm{x})e^{i\omega z+i\phi}$, where $e_{\mu}^0$ is the unit vector in time direction. Thus, Eq.~\eqref{eq:A5} gives
\begin{equation}
    \bm{A}_{5}(\bm{x}) = -\frac{4e}{\omega}\bm{e}_{3}f^{2}(\bm{x}),
\end{equation}
where $\bm{e}_3$ is the unit vector in the $z$ direction.

Therefore, if we choose $f(\bm{x})$ to be uniform, i.e, $f(\bm{x})=A_{\omega}$, we obtain a constant axial field,
\begin{equation}
    \bm{A}_{5}=-\frac{4eA_{\omega}^{2}}{\omega}\bm{e}_{3}.
\end{equation}
Unlike the vector field, in the finite-mass situation, one cannot gauge out the above constant axial field, so it can lead to physical output. In the next two sections, we will discuss this situation in detail.

Another interesting situation is $f(\bm{x})=A_{\omega}\exp(-r^{2}/(2\sigma^{2}))$, where $r=\sqrt{x^2+y^2}$, which means that the circular polarized wave is a Gaussian beam.
In this case, we find out that the following effective axial magnetic field occurs in the system:
\begin{equation}
    \bm{B}_{5}=\nabla\times\bm{A}_5=\frac{8eA_{\omega}^{2}}{\omega\sigma^{2}}e^{-\frac{r^{2}}{\sigma^{2}}}\left(\bm{x}\times\bm{e}_{z}\right).
\end{equation}
Interestingly, this $\bm{B}_{5}$ can induce the axial magnetic effect in the system after the Schwinger effect has produced some fermions and generate a charge current with density $\bm{j} = e\mu/(2\pi^2) \bm{B}_5$, where $\mu$ is the chemical potential of the fermions.
(See \cite{Gorbar:2016ygi, Landsteiner:2017hye, Ilan:2019lqk} for in-depth discussion of the axial magnetic effect.)
Moreover, since $\bm{B}_5$ is vortical, the charge current is vortical as well.
These vortical currents may have interesting implications to the evolution of the system.

To conclude this section, let us put some remarks on the regime of $\omega$ for the high-frequency effective theory to be valid.
According to current research, when a system obtains energy from high-frequency fields, the convergence condition and convergence speed of the Floquet-Magnus expansion become controversial, both in the context of absolute convergence and asymptotic convergence. See \cite{Bukov04032015} for a comprehensive discussion and \cite{Blanes:2008xlr, Goldman:2014xja, Managa2016} for relevant theoretical proofs.
To avoid the complexity, we only qualitatively estimate the regime of $\omega$ based on the condition $||\hat{H}_F^{(1)}||/||\hat{H}_F^{(0)}||\ll 1$.
From Eq.~\eqref{eq:HF-0}, when the field strength of the static electric field is small, $||\hat{H}_F^{(0)}||\sim m$; when the field strength is comparable to $eE_S\sim m^2$, $||\hat{H}_F^{(0)}||\sim a m^2$, where $a$ is the length scale of the static field $\varphi(\bm{x})$.
From Eq.~\eqref{eq:HF-1-interm}, $||\hat{H}_F^{(1)}||\sim e^2E_{\omega}^2/\omega^3$, where $E_{\omega} \sim \omega A_{\omega}$ is the electric field strength of the high-frequency field.
Thus, when the static electric field is weak, $||\hat{H}_F^{(1)}||/||\hat{H}_F^{(0)}|| \ll 1$ becomes
\begin{equation}
    \left(
        \frac{e E_{\omega}}{m \omega}
    \right)^2 \frac{m}{\omega} = \gamma_K^2 \frac{m}{\omega} \ll 1.
\end{equation}
When the static electric field strength is comparable to $E_S$, the condition becomes
\begin{equation}
    \left(
        \frac{e E_{\omega}}{m \omega}
    \right)^2 \frac{1}{a\omega} = \gamma_K^2 \frac{1}{a\omega} \ll 1.
\end{equation}
Here, $\gamma_K=eE_{\omega}/(m\omega)$ is the Keldysh parameter of the high-frequency field, which is frequently used in strong-field QED as a measure of perturbativeness.
See \cite{Huang:2019uhf, DiPiazza:2015xva, Taya:2014taa} for discussions on it.
The above results show that the validity condition of the high-frequency effective theory developed here is closely related to that of the perturbative theory.
On the other hand, when $\omega$ is large, the high-frequency effective theory converges faster in the $\gamma_K\sim 1$ case.

\section{Numerical method}
\label{sec:setup}
Let us discuss the numerical approach to obtaining the number of fermions in each state.

For the external field in our setup, we follow the discussion in the previous section and choose $A^{0}(x) = \varphi(\bm{x})$, $\bm{A}(x) = (\bm{e}_{1}-i\bm{e}_{2}) A_{\omega} e^{i\omega (z-t)} + \mathrm{c.c.}$
We fix the static field as the following periodic one:
\begin{equation}
    e\varphi(\bm{x})=\frac{eV_{0}}{\cosh(z/a)},\,-L/2<z<L/2,
\end{equation}
\begin{equation}
    \varphi(z+L)=\varphi(z),
\end{equation}
where $eV_0$ is a parameter with the mass dimension $+1$, and $L$ and $a$ are two parameters with mass dimension $-1$.
We require $L$ to be sufficiently large such that $L^{-1}$ is small compared with the momentum scale of interest.
In addition, we require $L$ to be integer times of $2\pi/\omega$.

With the above external field, $\hat{H}_F$ and $\hat{K}(t)$ become uniform in $x$ and $y$ directions and periodic in the $z$ direction with period $L$.
In $-L/2<z<L/2$ and up to the first order, they are
\begin{align}
    &\hat{H}_F = -i\gamma^{0}\gamma^{3}\partial_{z}+\gamma^{0}\bm{\gamma}\cdot\bm{p}_T+m\gamma^{0}\nl
    +\frac{eV_{0}}{\cosh(z/a)}+\frac{4e^{2}A_{\omega}^{2}}{\omega}\gamma^{0}\gamma^{3}\gamma^{5},
    \label{eq:HF-final}
\end{align}
\begin{equation}
    \hat{K}(t) = \frac{2eA_{\omega}}{\omega}\gamma^{0}\left(\gamma^{1}\sin(\omega z-\omega t)-\gamma^{2}\cos(\omega z-\omega t)\right),
\end{equation}
where $\bm{p}_T=(p_x,p_y,0)^T$ is the momentum of the fermions in $x$ and $y$ directions.
Later we refer to it as the transverse momentum.

In Sec.~\ref{sec:eff-theory}, we show that the eigenmodes $\ket{u^{\gamma}_F}$, $\ket{v^{\gamma}_F}$ satisfy $\hat{H}_{F} \ket{u_{F}^{\gamma}} = \epsilon_{F}^{\gamma} \ket{u_{F}^{\gamma}}$, $\hat{H}_{F} \ket{v_{F}^{\gamma}} = -\bar{\epsilon}_{F}^{\gamma} \ket{v_{F}^{\gamma}}$.
Since $\hat{H}_{F}$ is periodic in the $z$ direction and $L$ is large, we can choose $u^{\gamma}_F(\bm{x})$ and $v^{\gamma}_F(\bm{x})$ that obey either the periodic or antiperiodic boundary condition in the $z$ direction. 
Among the two boundary conditions, the advantage of the antiperiodic one is the ability to remove the possible singularity of the Green's function at zero momentum (see \cite{Fukushima:2015tza} for a discussion).
In our calculation, however, we work on the finite-mass situation where the singularity does not occur, so we choose the periodic boundary condition.
Hence, we consider the following ansatz of the eigenmodes:
\begin{equation}
    u^{\gamma}_F(\bm{x})=u^{\bm{p}_T,k}_F(z)e^{i\bm{p}_T\cdot\bm{x}},
\end{equation}
\begin{equation}
    v^{\gamma}_F(\bm{x})=v^{-\bm{p}_T,k}_F(z)e^{i\bm{p}_T\cdot\bm{x}}.
\end{equation}
Here, $u^{\bm{p}_T,k}_F(z)=u^{\bm{p}_T,k}_F(z+L)$ and $v^{-\bm{p}_T,k}_F(z)=v^{-\bm{p}_T,k}_F(z+L)$.
In these equations, we identify the abstract index of the eigenmodes $\gamma$ to $\{\bm{p}_T, k\}$ for positive energy states and $\{-\bm{p}_T, k\}$ for negative energy states, where $\bm{p}_T$ is the transverse momentum that is continuous, and $k$ is an index corresponding to the $z$ momentum and spin, which takes discrete values thanks to the periodic boundary condition.
Since $\hat{H}_F$ is time independent, we do not need to include an ansatz of Bloch form in the time direction.

In order to solve $u^{\bm{p}_T,k}_F(z)$ and $v^{-\bm{p}_T,k}_F(z)$ from the static $\hat{H}_{F}$, one may consider (1) the WKB approximation, (2) discretizing $\hat{H}_F$ and solving the eigenvalue equations on lattices. 
To use the WKB approximation, one needs to compute the classical trajectories of the fermions under $\hat{H}_{F}$.
(See \cite{DiPiazza:2013vra} for the WKB approximation to diagonalize the Dirac Hamiltonian.)
However, since the $\hat{H}_{F}$ presented above involves both massive fermions and an axial field, it is not yet clear how to define the classical trajectories.
Therefore, we choose to discretize $\hat{H}_F$ and solve the eigenvalue equations on lattices.

In the lattice approach, the most crucial step is to eliminate the fermion doublers, which are fictitious eigenmodes of the discretized Dirac Hamiltonian that arise when one tries to naively discretize the differential operator as $\partial_{\mu} f(x) \to (f(x+\Delta x^{\mu}/2)-f(x-\Delta x^{\mu}/2))/\Delta x^{\mu}$, where $\Delta x^{\mu}$ is the lattice spacing in the $\mu$ direction.
(See \cite{Fukushima:2015tza} for more discussion of doublers.)
For example, if we choose $V_0=0$, $A_{\omega}=0$ and discretize $\hat{H}_F$ defined in Eq.~\eqref{eq:HF-final} naively, we will find that for each $\epsilon^{\bm{p}_T,k}$ (or $\bar{\epsilon}^{-\bm{p}_T,k}$), $u^{\bm{p}_T,k}_F(z)$ [or $v^{-\bm{p}_T,k}_F(z)$] is of eightfold degeneracy.
However, in the continuous limit, the correct answer should be fourfold degeneracy, two from the direction of motion and two from spin.
To eliminate the doublers, one strategy is to introduce extra terms to the discretized Hamiltonian that explicitly break some symmetries of the Hamiltonian (but recover the symmetries in the continuous limit).
For this purpose, we add the following Wilson term $\hat{H}_W$ to the discretized Hamiltonian:
\begin{equation}
    \hat{H}_{W} f(z)=-\frac{
        f(z+\Delta z)+f(z-\Delta z)-2f(z)
    }{\Delta z}.
\end{equation}
This term approaches $-\partial_z^2f(z)\Delta z$ for small $\Delta z$ and has no influence in the continuous limit.
In the finite-$\Delta z$ case, it eliminates the doublers and allows us to obtain $u^{\bm{p}_T,k}_F(z)$ and $v^{-\bm{p}_T,k}_F(z)$.

When $V_0=0$, $A_{\omega}=0$, the $\hat{H}_F$ in Eq.~\eqref{eq:HF-final} equals $\hat{H}(t_{\text{in}})$.
As a result, one can follow the same procedure presented above and calculate $\ket{u^{\bm{p}_T,k}(t_{\text{in}})}$, $\ket{v^{-\bm{p}_T,k}(t_{\text{in}})}$ by discretizing $\hat{H}(t_{\text{in}})$ and solving the eigenvalue equations on lattices.

\begin{table}
    \caption{Parameters in the numerical calculation.
    $\omega$ is the frequency of the high-frequency field. 
    When performing the numerical calculation, we normalize all quantities with mass dimension $+1$ by $\omega$, and those with mass dimension $-1$ by $\omega^{-1}$.}
    \label{tab:parameters}
    \begin{ruledtabular}
        \begin{tabular}{l l}
        Parameter name & Value\\
        Fermion mass ($m$)    &   $10.00$ $\omega$\\
        Shape parameter of the static field ($a$) &   $0.50\pi$ $\omega^{-1}$\\
        Magnitude of the static field ($eV_0$)   &   $-2am^2$\\
        Magnitude of the high-frequency field ($eA_{\omega}$) &   $0.00$---$2.00$ $\omega$\\
        Period in $z$ direction ($L$) &   $8.00\pi$ $\omega^{-1}$\\
        Number of lattices in $z$ direction ($N_z$) &   $500$\\
        Transverse momentum ($p_{T}$)    &   $0.00$---$100.00$ $\omega$\\
        End time of the short-time calculation ($t_0$)  &   $\frac{(N_z-1)/L}{|eV_0|/(2a)}$\\
        \end{tabular}
    \end{ruledtabular}
\end{table}

Once we calculate both $\ket{u^{\bm{p}_T,k}_F}$, $\ket{v^{-\bm{p}_T,k}_F}$ and $\ket{u^{\bm{p}_T,k}(t_{\text{in}})}$, $\ket{v^{-\bm{p}_T,k}(t_{\text{in}})}$, we substitute them into Eqs.~\eqref{eq:n-t}, \eqref{eq:S} and Eqs.~\eqref{eq:Splus}, \eqref{eq:Sminus}, \eqref{eq:n-long-time} to compute the $n^{\bm{p}_T,k}(t)$ and $n^{\bm{p}_T,k}$ in the long-time limit, respectively.
In the current work, we focus on the energy spectrum of the fermions number, so we sum up the degenerate states of fermions with the same energy, and define
\begin{equation}
    n^{\bm{p}_T}(t;\epsilon) = \sum_{k, \epsilon=\epsilon^{\bm{p}_T,k}} n^{\bm{p}_T,k}(t),
\end{equation}
\begin{equation}
    n^{\bm{p}_T}(\epsilon) = \sum_{k, \epsilon=\epsilon^{\bm{p}_T,k}} n^{\bm{p}_T,k},
\end{equation}
In the next section, we compute $n^{\bm{p}_T}(t;\epsilon)$ and $n^{\bm{p}_T}(\epsilon)$ with parameters illustrated in Table~\ref{tab:parameters}.

\section{Results and discussions}
\label{sec:results}

In this section, we present and discuss the numerical results, focusing on the number of fermions with energy $\epsilon$ and transverse momentum $\bm{p}_T$.

\begin{figure}
    \centering
    \includegraphics[width=\linewidth]{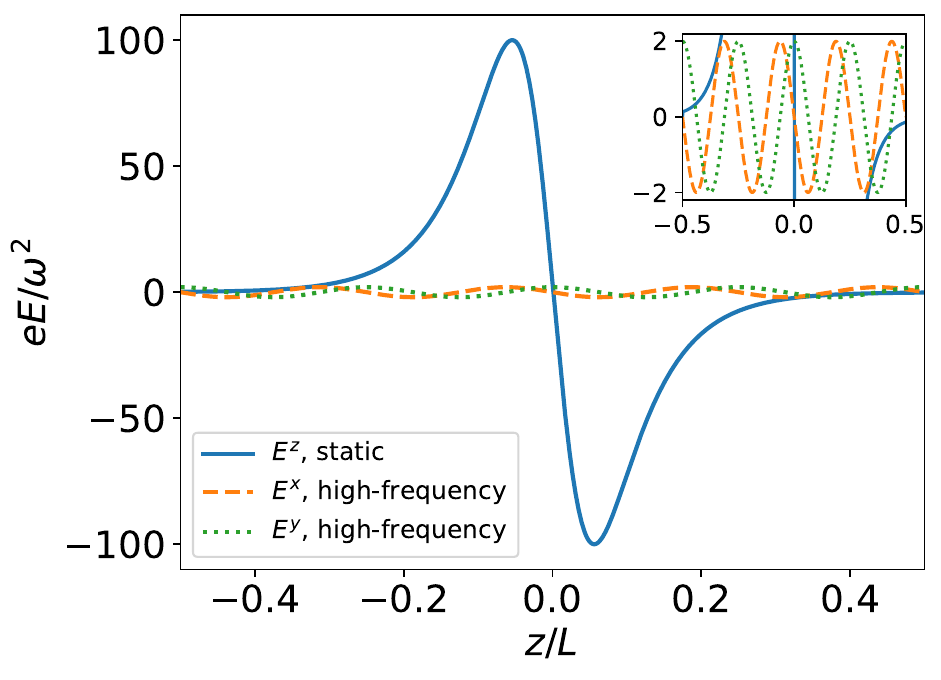}
    \caption{External electric strength in one period of $z$ at $t=0$. The $z$ component is the static field. The $x$ and $y$ components are the high-frequency fields. Here, we choose $eA_{\omega} = 1.00\omega$. All other parameters are as shown on Table~\ref{tab:parameters}.}
    \label{fig:external-fields}
\end{figure}

We illustrate the static and high-frequency external fields at $t=0$ on Fig.~\ref{fig:external-fields}.
According to this figure, the magnitude of the high-frequency electric field is $E^{x}\sim E^{y} \sim \omega^2$. Hence, $\gamma_K \sim 0.1$ and the result from the high-frequency effective theory is valid.
In addition, Fig.~\ref{fig:external-fields} shows that at every $z$, the magnitude of the static electric field is below the Schwinger threshold $E_{S}\sim m^2=100.00\omega^2$, and the magnitude of the high-frequency electric field is far smaller than the threshold.
Without being dynamically assisted, the Schwinger effect in this setup is almost negligible.
The perturbative particle pair production is also negligible since $m \gg \omega$.
Therefore, all particle production in the system comes from the dynamically assisted Schwinger effect.

\begin{figure*}
    \begin{subfigure}{0.49\linewidth}
        \includegraphics[width=\linewidth]{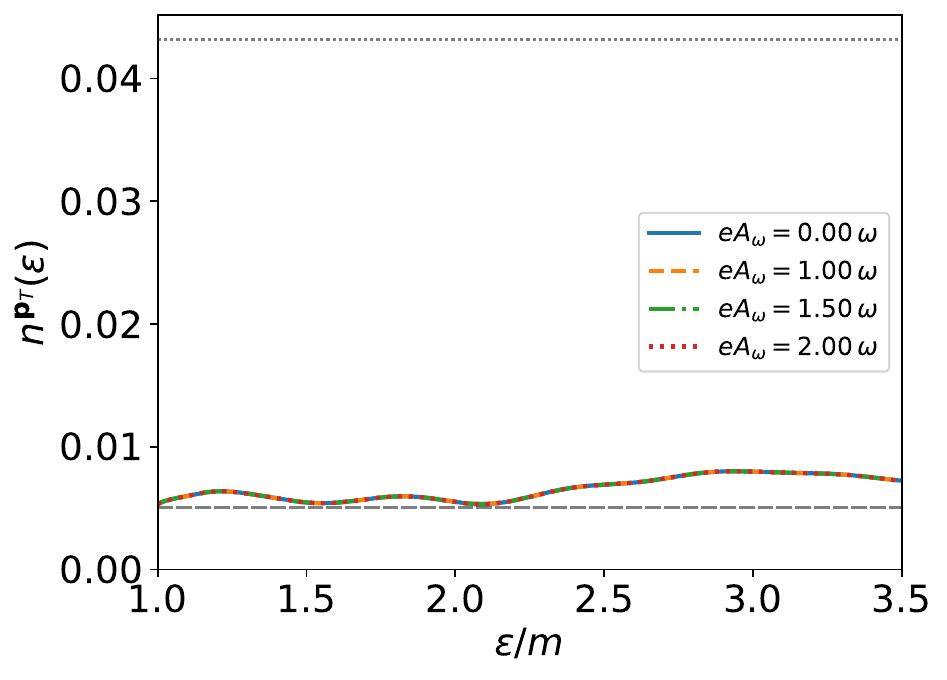}
        \caption{}
    \end{subfigure}
    \begin{subfigure}{0.49\linewidth}
        \includegraphics[width=\linewidth]{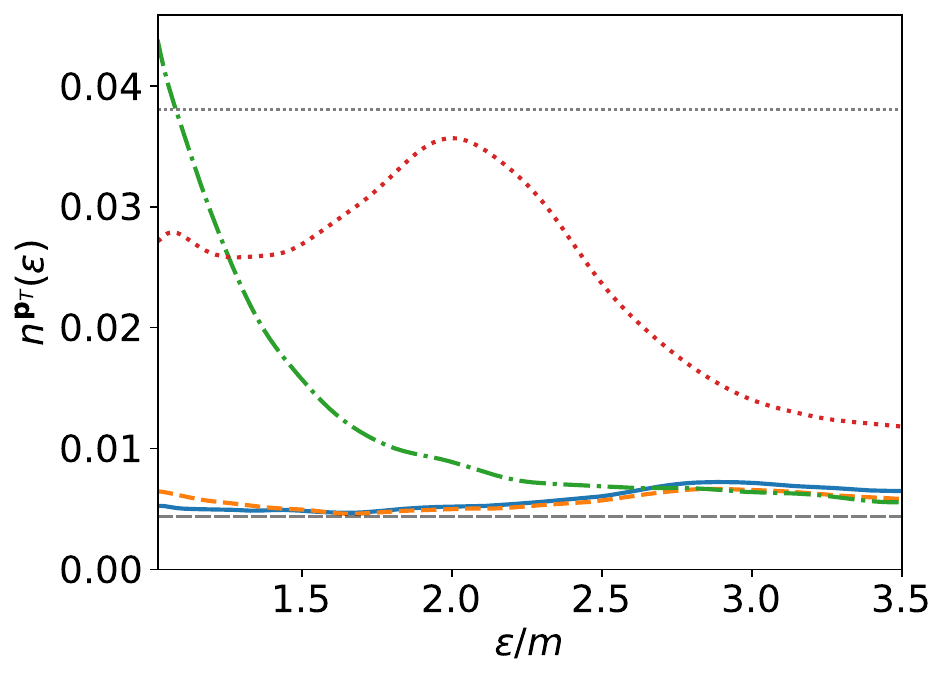}
        \caption{}
    \end{subfigure}
    \begin{subfigure}{0.49\linewidth}
        \includegraphics[width=\linewidth]{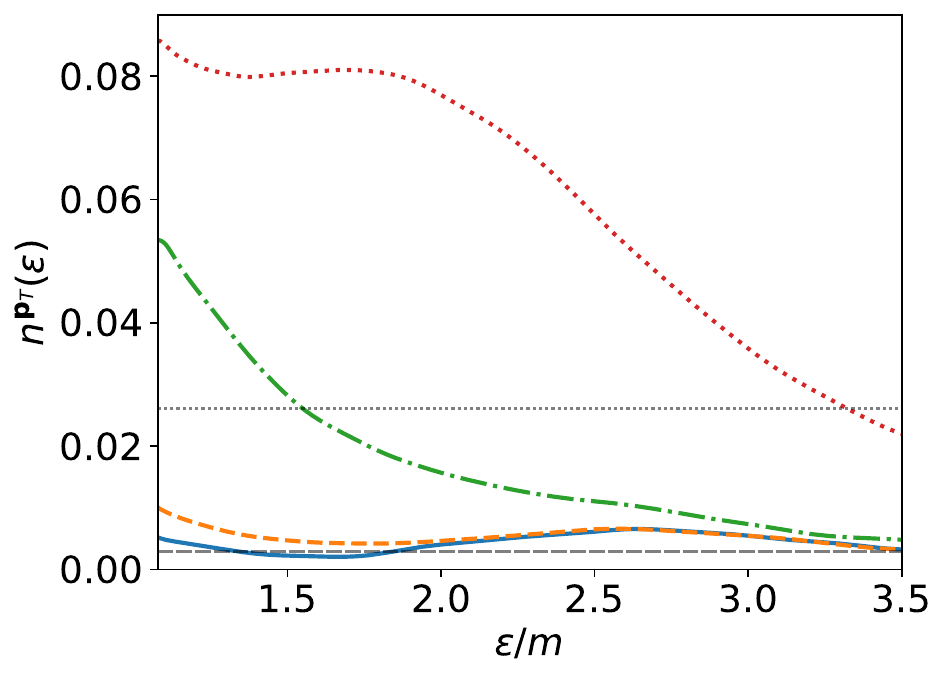}
        \caption{}
    \end{subfigure}
    \begin{subfigure}{0.49\linewidth}
        \includegraphics[width=\linewidth]{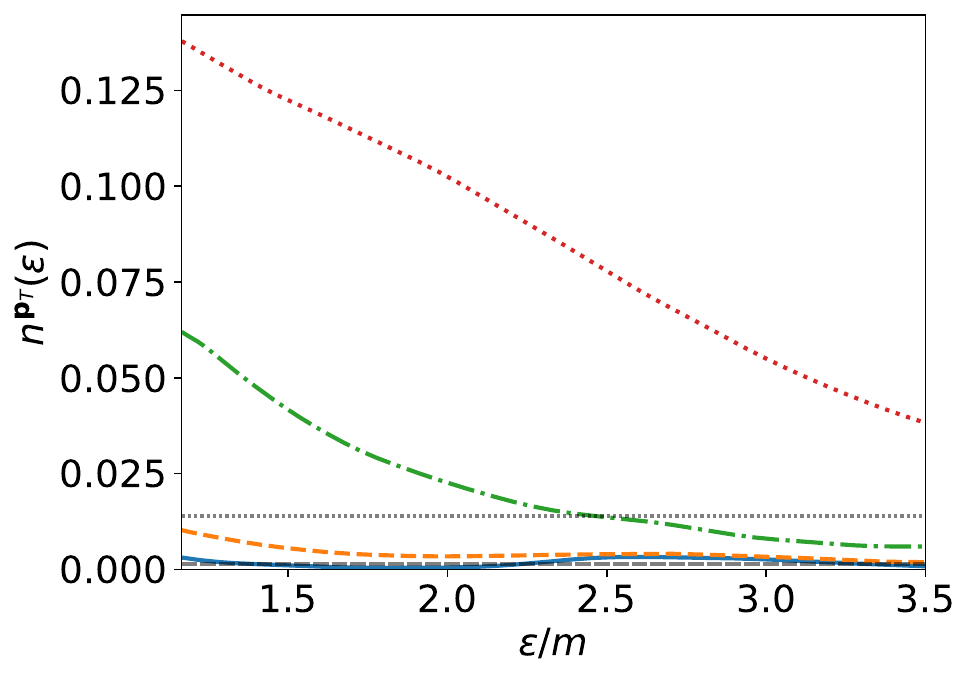}
        \caption{}
    \end{subfigure}
    \caption{Number of the positive-energy fermions with respect to the energy $\epsilon$ in the long-time limit.
    The values of the transverse momentum are (a) $p_T=0.00$, (b) $p_T=2.00\omega$, (c) $p_T=4.00\omega$, (d) $p_T=6.00\omega$, respectively.
    The dashed horizontal line and the dotted one represent the constant-field estimation and the threshold value of the fermion number, respectively.}
    \label{fig:long-time}
\end{figure*}

\begin{figure}
    \includegraphics[width=\linewidth]{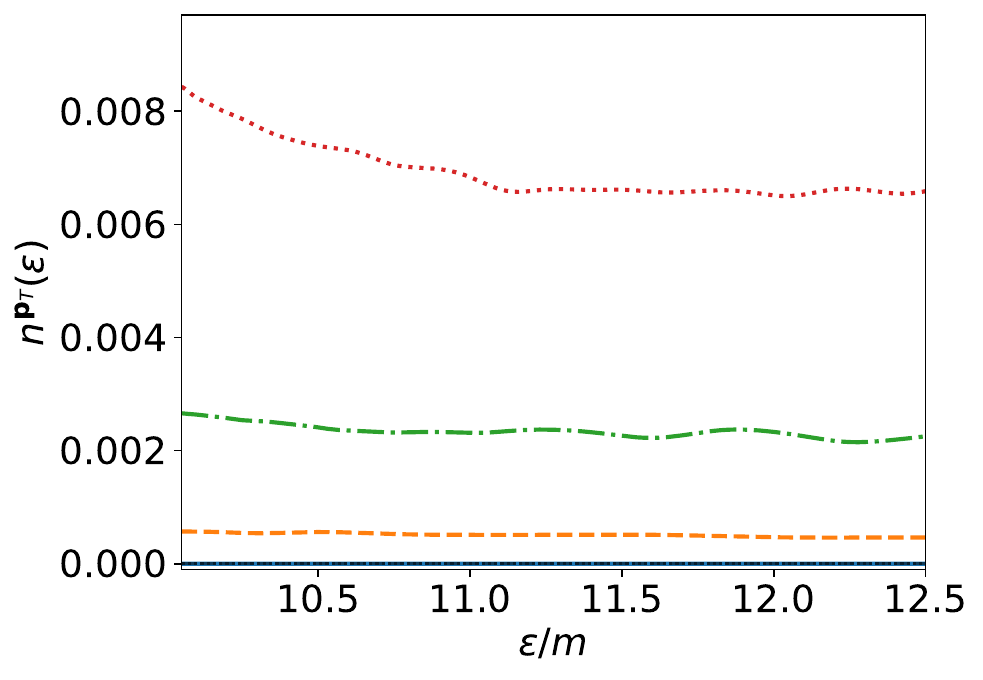}
    \caption{Number of the positive-energy fermions with respect to the energy $\epsilon$ in the long-time limit.
    The value of the transverse momentum is $p_T=100.00\omega$.
    The conventions of different lines are the same as Fig.~\ref{fig:long-time}.
    }
    \label{fig:long-time-large-pT}
\end{figure}

In the long-time limit, we plot $n^{\bm{p}_T}(\epsilon)$ with $\bm{p}_T=0.00, 2.00\omega, 4.00\omega, 6.00\omega$ on Fig.~\ref{fig:long-time}, and $n^{\bm{p}_T}(\epsilon)$ with $\bm{p}_T=100.00\omega$ on Fig.~\ref{fig:long-time-large-pT}.
Hence, the two figures show the particle production in directions close to the $z$ axis and directions almost transverse to it, receptively.
To demonstrate the influence of the axial field, we also introduce two axillary quantities.
The first is the number of fermions produced by a constant electric field $\bm{E}=\bm{e}_3 |\frac{1}{L}\int_{-L/2}^{L/2} dz \, E^z(z)|$ with $E^z(z)$ as shown on Fig.~\ref{fig:external-fields}:
\begin{equation}
    n^{\bm{p}_T}_0 = e^{-\frac{\pi(m^2+\bm{p}_T^2)}{\left|e\frac{1}{L}\int_{-L/2}^{L/2} dz \, E^z(z)\right|}}.
\end{equation}
One can view this quantity as a rough estimation of the number of fermions produced solely by the static field, which is independent of the theoretical model in this work.
The second one is the number of fermions produced by a constant electric field with field strength at the Schwinger threshold:
\begin{equation}
    n^{\bm{p}_T}_S = e^{-\frac{\pi(m^2+\bm{p}_T^2)}{m^2}}.
\end{equation}
We use this quantity to estimate the threshold value of the fermion number that is experimentally observable.
Both of these two quantities appear on Figs.~\ref{fig:long-time} and \ref{fig:long-time-large-pT} as the horizontal lines.
From then on, we refer to them as the constant-field estimation of the fermion number and the threshold value of the fermion number, respectively.

From Figs.~\ref{fig:long-time} and \ref{fig:long-time-large-pT}, we find that
\begin{itemize}
    \item When the axial field vanishes ($eA_{\omega}=0$), the number of particle stays close to $n^{\bm{p}_T}_0$.
    This result, on the one hand, verifies the validity of our numerical method and, on the other hand, shows that unless the peak value of electric field strength crosses the Schwinger threshold, modifying the spatial distribution of the field does not significantly enhance the particle production.
    Hence, all increases in the number of fermions produced are most likely to come from the axial field.
    \item When $A_{\omega}\neq 0$, even a weak axial field can strongly enhance the particle production.
    In our case, the maximum magnitude of the effective axial field is $eA_5 = 16.00\omega$, significantly smaller than the maximum magnitude of the static field, $eV_0\approx 314.00\omega$.
    However, before applying the axial field, the number of fermions is close to $n^{\bm{p}_T}_0$, far below $n^{\bm{p}_T}_S$;
    after applying, it can easily approach or exceed $n^{\bm{p}_T}_S$, becoming experimentally observable.
    \item The axial field enhances the production of low-energy fermions more than the production of high-energy ones.
    One can understand this behavior in the following way.
    In the high-energy limit, the fermions are effectively massless.
    At this situation, the left-handed and right-handed sectors are decoupled, so $\bm{A}_5 = -(4e A_{\omega}^2)/\omega \bm{e}_{3}$ is identical to $\bm{A} = -(4e A_{\omega}^2)/\omega \bm{e}_{3}$ acting on the left-handed sector and $\bm{A} = (4e A_{\omega}^2)/\omega \bm{e}_{3}$ acting on the right-handed sector.
    These constant gauge fields can be eliminated by gauge transformation, so do not have any physical effect.
    Hence, the axial-field enhancement of the particle production becomes weak at high energy.
    \item The axial field does not enhance the production of fermions moving in the $z$ direction.
    This phenomenon originates from the fact that the effective axial field is induced by a circular polarized plane wave propagating in $z$ direction.
    For this field, both the electric and magnetic fields are in the transverse direction.
    Hence, fermions moving in the $z$ direction do not have any energy exchange with the high-frequency field.
    \item Figure \ref{fig:long-time} shows that, for fermions moving in directions close to the $z$ axis, increasing $\bm{p}_T$ leads to the increase of the number of fermions; on the other hand, Fig.~\ref{fig:long-time-large-pT} shows that, for the fermions moving in directions almost transverse to the $z$ axis, the number of fermions decreases to a negligibly small value.
    The behavior suggests that $n^{\bm{p}_T}(\epsilon)$ has a nonmonotonic dependence on the scatter angle $\theta = \arcsin (p_T/\sqrt{\epsilon^2-m^2})$, and reaches the maximum value in some intermediate values of $\theta$ ($\theta \neq 0, \pi/2$).
    This observation is in agreement with the conclusion of Huang \textit{et al.} in \cite{Huang:2019uhf}, which studied the dynamically assisted Schwinger effect under a uniform high-frequency field that is circular polarized.
    This angular dependence is related to the spin-dependent coupling between the fermions and the axial fields.
\end{itemize}

\begin{figure}
    \includegraphics[width=\linewidth]{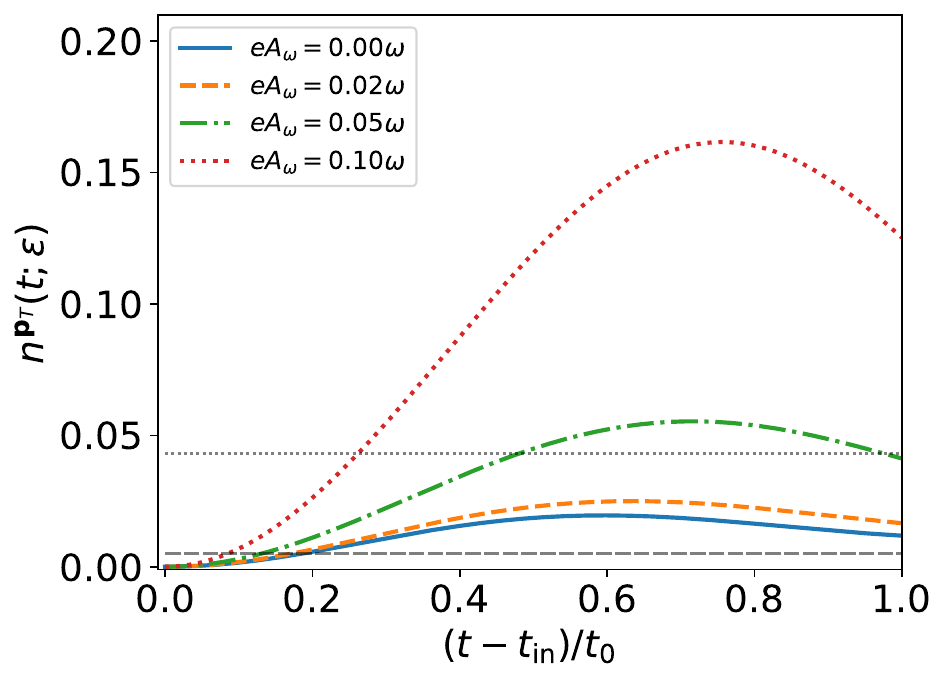}
    \caption{Number of the positive-energy fermions with respect to time.
    The energy of the fermion is $\epsilon=m$. 
    The transverse momentum is $\bm{p}_T=0$.
    The dashed horizontal line and the dotted one represent the constant-field estimation and the threshold value of the fermion number, respectively.}
    \label{fig:short-t-dep}
\end{figure}

On the short timescale, $n^{\bm{p}_T}(t;\epsilon)$ shows different behaviors.
To illustrate this, we plot $n^{\bm{p}_T}(t;\epsilon)$ with respect to $t$ at $\epsilon=m$, $\bm{p}_T=0$ in Fig.~\ref{fig:short-t-dep}.
We choose these values of $\epsilon$ and $\bm{p}_T$ because in the Schwinger effect, fermion pairs are first generated with zero energy and then accelerated to high energy by the electric field, so $n^{\bm{p}_T=0}(t;\epsilon=m)$ measures the number of newly produced fermions.

From this figure, we observe that
\begin{itemize}
    \item When there is only the static electric field, the number of fermions first increases steadily from zero and reaches a peak value that is higher than $n^{\bm{p}_T}_0$;
    after that, it slowly decreases to approach the $n^{\bm{p}_T}_0$.
    This nonmonotonic behavior is due to a sudden switch on of the electric field at $t=t_{\text{in}}$, which introduces high-frequency fields that enhance the particle production;
    when the electric field becomes constant at later time, the fermion number relaxes to the constant-field estimation.
    \item When the axial field presents, the increase and fall in the number of fermions becomes more drastic, inducing a significantly larger maximum fermion number.
    Interestingly, on the short timescale, the particle production enhancement is much stronger than that on the long timescale.
    On the long timescale, one needs $eA_{\omega}\sim 1.00\omega$ to generate observable fermion number; but on the short timescale, one needs merely $eA_{\omega}\sim 0.10\omega$.
    This difference is the consequence of the initial kick from the axial field, embedded in the $\exp(i\hat{K}(t_{\text{in}}))$ and $\exp(-i\hat{K}(t))$ factors that vanish on the long timescale.
\end{itemize}
According to the above observations, if one can use short pulses of axial fields with proper length in the dynamically assisted Schwinger effect, one may acquire stronger particle production enhancement than that from the constant axial field.

\section{Summary}
\label{sec:summary}

In this work, we study the role of spatial axial fields in the dynamically assisted Schwinger effect in the context of high-frequency effective theory.

First, we establish the high-frequency effective theory of the dynamically assisted Schwinger effect based on the Floquet-Magnus expansion and derive the number of fermions in each state, both the time-dependent and long-time results.
Then, we study a field configuration consisting of a static electric field and a high-frequency circular polarized wave, which is easy to implement in experiments, and show that this configuration induces a spatial axial field in the high-frequency effective theory.
Finally, we develop the numerical approach to obtain the number of fermions produced with specific energy and transverse momentum, and discuss how it behaves across different timescales.

From these discussions, we discover that
\begin{enumerate}
    \item The axial field enhances the particle production of the dynamically assisted Schwinger effect both on the long and short timescale.
    \item On the long timescale, the axial-field enhancement is particularly strong for large-mass fermions.
    For fermions moving in the direction parallel to the axial field, however, the enhancement diminishes.
    \item On the short timescale, the initial kick from the kick operator of the axial field plays an important role in enhancing the particle production.
    It induces the rapid rise and fall of the fermion number, with a peak value that is larger than the fermion number on the long timescale.
\end{enumerate}

In conclusion, we show that axial fields can easily occur in the dynamically assisted Schwinger effect as the effective fields of the high-frequency fields and significantly enhance the phenomenon, offering both theoretical insights and useful tools for the experimental implementation of the Schwinger effect.

Future extensions of this work include: (1) solving the eigenmodes $\ket{u_{F}^{\alpha}}$, $\ket{v_{F}^{\alpha}}$, $\ket{u^{\alpha}_{\text{in}}(t_{\text{in}})}$, $\ket{v^{\alpha}_{\text{in}}(t_{\text{in}})}$ analytically based on the WKB approximation; (2) studying the spin-dependent coupling between the axial field and the fermions and the induced angular dependence of particle production; (3) comparing the results on the short timescale with the results from kinetic theories; and (4) order-by-order comparison between the high-frequency effective theory and perturbative QED.

\appendix
\section{DERIVING EQ.\eqref{eq:HF-A5}}
\label{app:HF-A5}
Let us prove
\begin{equation}
    [\gamma^{0}\gamma^{\mu},\gamma^{0}\gamma^{\nu}] \tilde{A}_{\mu} \tilde{A}_{\nu}^{*} = 2i\epsilon^{ijk}\gamma^{0}\gamma^{k}\gamma^{5} \tilde{A}_i \tilde{A}^*_j.
\end{equation}

To start with, we know $(\gamma^0)^2=I$, so
\begin{equation}
    [\gamma^{0}\gamma^{\mu},\gamma^{0}\gamma^{0}] \tilde{A}_{\mu} \tilde{A}_{0}^{*}
    =[\gamma^{0}\gamma^{0},\gamma^{0}\gamma^{\mu}] \tilde{A}_{0} \tilde{A}_{\mu}^{*}
    =0.
\end{equation}
This yields
\begin{equation}
    [\gamma^{0}\gamma^{\mu},\gamma^{0}\gamma^{\nu}] \tilde{A}_{\mu}\tilde{A}_{\nu}^{*}=[\gamma^{0}\gamma^{i},\gamma^{0}\gamma^{j}] \tilde{A}_{i}\tilde{A}_{j}^{*}.
    \label{eq:AiAj}
\end{equation}

Next, we discuss $[\gamma^{0}\gamma^{i},\gamma^{0}\gamma^{j}]$.
When $i=j$, $[\gamma^{0}\gamma^{i},\gamma^{0}\gamma^{j}]=0$.
Hence, we focus on the $i\neq j$ situation.
Without loss of generality, we choose $i=1$, $j=2$.
This yields
\begin{align}
    &[\gamma^{0}\gamma^{1},\gamma^{0}\gamma^{2}]=[\gamma^{1},\gamma^{2}]+\gamma^{0}[\gamma^{1},\gamma^{0}]\gamma^{2}+\gamma^{0}[\gamma^{0},\gamma^{2}]\gamma^{1}\nl
    =-2\gamma^{1}\gamma^{2}.
\end{align}
At the same time, $\gamma^{5}=i\gamma^{0}\gamma^{1}\gamma^{2}\gamma^{3}$, so
\begin{equation}
    \gamma^{0}\gamma^{3}\gamma^{5}=i\gamma^{1}\gamma^{2}=\frac{1}{2i}[\gamma^{0}\gamma^{1},\gamma^{0}\gamma^{2}].
\end{equation}
Using completely the same method, one can show
\begin{equation}
    \gamma^{0}\gamma^{1}\gamma^{5}=\frac{1}{2i}[\gamma^{0}\gamma^{2},\gamma^{0}\gamma^{3}],
\end{equation}
\begin{equation}
    \gamma^{0}\gamma^{2}\gamma^{5}=\frac{1}{2i}[\gamma^{0}\gamma^{3},\gamma^{0}\gamma^{1}].
\end{equation}
Therefore,
\begin{equation}
    \gamma^{0}\gamma^{k}\gamma^{5}=\frac{1}{2i}\frac{\epsilon^{ijk}}{2}[\gamma^{0}\gamma^{i},\gamma^{0}\gamma^{j}],
\end{equation}
\begin{equation}
    [\gamma^{0}\gamma^{i},\gamma^{0}\gamma^{j}] =2i\epsilon^{ijk}\gamma^{0}\gamma^{k}\gamma^{5}.
\end{equation}

Finally, substituting the above result back to Eq.~\eqref{eq:AiAj}, we prove
\begin{equation}
    [\gamma^{0}\gamma^{\mu},\gamma^{0}\gamma^{\nu}] \tilde{A}_{\mu}\tilde{A}_{\nu}^{*} = 2i\epsilon^{ijk} \gamma^{0}\gamma^{k}\gamma^{5} \tilde{A}_i\tilde{A}^*_j
\end{equation}

This relation shows that Eq.~\eqref{eq:HF-1-interm} is identical to Eq.~\eqref{eq:HF-A5}.
This finishes the derivation.

\begin{acknowledgments}
    This research is under the support of KAKENHI Grant No. JP23K25864, and in part used the computational resources provided by Multidisciplinary Cooperative Research Program in the Center for Computational Sciences, University of Tsukuba.
    We thank Professor Takashi Nakatsukasa of University of Tsukuba for constructive discussions in the proposal phase of the project, and Professor Kenji Fukushima of the University of Tokyo for discussions on the numerical method.
\end{acknowledgments}

\bibliographystyle{apsrev4-2}
\bibliography{refs.bib}

\begin{thebibliography}{59}%
\makeatletter
\providecommand \@ifxundefined [1]{%
 \@ifx{#1\undefined}
}%
\providecommand \@ifnum [1]{%
 \ifnum #1\expandafter \@firstoftwo
 \else \expandafter \@secondoftwo
 \fi
}%
\providecommand \@ifx [1]{%
 \ifx #1\expandafter \@firstoftwo
 \else \expandafter \@secondoftwo
 \fi
}%
\providecommand \natexlab [1]{#1}%
\providecommand \enquote  [1]{``#1''}%
\providecommand \bibnamefont  [1]{#1}%
\providecommand \bibfnamefont [1]{#1}%
\providecommand \citenamefont [1]{#1}%
\providecommand \href@noop [0]{\@secondoftwo}%
\providecommand \href [0]{\begingroup \@sanitize@url \@href}%
\providecommand \@href[1]{\@@startlink{#1}\@@href}%
\providecommand \@@href[1]{\endgroup#1\@@endlink}%
\providecommand \@sanitize@url [0]{\catcode `\\12\catcode `\$12\catcode `\&12\catcode `\#12\catcode `\^12\catcode `\_12\catcode `\%12\relax}%
\providecommand \@@startlink[1]{}%
\providecommand \@@endlink[0]{}%
\providecommand \url  [0]{\begingroup\@sanitize@url \@url }%
\providecommand \@url [1]{\endgroup\@href {#1}{\urlprefix }}%
\providecommand \urlprefix  [0]{URL }%
\providecommand \Eprint [0]{\href }%
\providecommand \doibase [0]{https://doi.org/}%
\providecommand \selectlanguage [0]{\@gobble}%
\providecommand \bibinfo  [0]{\@secondoftwo}%
\providecommand \bibfield  [0]{\@secondoftwo}%
\providecommand \translation [1]{[#1]}%
\providecommand \BibitemOpen [0]{}%
\providecommand \bibitemStop [0]{}%
\providecommand \bibitemNoStop [0]{.\EOS\space}%
\providecommand \EOS [0]{\spacefactor3000\relax}%
\providecommand \BibitemShut  [1]{\csname bibitem#1\endcsname}%
\let\auto@bib@innerbib\@empty
\bibitem [{\citenamefont {Schwinger}(1951)}]{Schwinger:1951nm}%
  \BibitemOpen
  \bibfield  {author} {\bibinfo {author} {\bibfnamefont {J.~S.}\ \bibnamefont {Schwinger}},\ }\href {https://doi.org/10.1103/PhysRev.82.664} {\bibfield  {journal} {\bibinfo  {journal} {Phys. Rev.}\ }\textbf {\bibinfo {volume} {82}},\ \bibinfo {pages} {664} (\bibinfo {year} {1951})}\BibitemShut {NoStop}%
\bibitem [{\citenamefont {Gelis}\ and\ \citenamefont {Tanji}(2016)}]{Gelis:2015kya}%
  \BibitemOpen
  \bibfield  {author} {\bibinfo {author} {\bibfnamefont {F.}~\bibnamefont {Gelis}}\ and\ \bibinfo {author} {\bibfnamefont {N.}~\bibnamefont {Tanji}},\ }\href {https://doi.org/10.1016/j.ppnp.2015.11.001} {\bibfield  {journal} {\bibinfo  {journal} {Prog. Part. Nucl. Phys.}\ }\textbf {\bibinfo {volume} {87}},\ \bibinfo {pages} {1} (\bibinfo {year} {2016})},\ \Eprint {https://arxiv.org/abs/1510.05451} {arXiv:1510.05451 [hep-ph]} \BibitemShut {NoStop}%
\bibitem [{\citenamefont {Torgrimsson}(2019)}]{Torgrimsson:2018xdf}%
  \BibitemOpen
  \bibfield  {author} {\bibinfo {author} {\bibfnamefont {G.}~\bibnamefont {Torgrimsson}},\ }\href {https://doi.org/10.1103/PhysRevD.99.096002} {\bibfield  {journal} {\bibinfo  {journal} {Phys. Rev. D}\ }\textbf {\bibinfo {volume} {99}},\ \bibinfo {pages} {096002} (\bibinfo {year} {2019})},\ \Eprint {https://arxiv.org/abs/1812.04607} {arXiv:1812.04607 [hep-ph]} \BibitemShut {NoStop}%
\bibitem [{\citenamefont {Taya}\ \emph {et~al.}(2021)\citenamefont {Taya}, \citenamefont {Fujimori}, \citenamefont {Misumi}, \citenamefont {Nitta},\ and\ \citenamefont {Sakai}}]{Taya:2020dco}%
  \BibitemOpen
  \bibfield  {author} {\bibinfo {author} {\bibfnamefont {H.}~\bibnamefont {Taya}}, \bibinfo {author} {\bibfnamefont {T.}~\bibnamefont {Fujimori}}, \bibinfo {author} {\bibfnamefont {T.}~\bibnamefont {Misumi}}, \bibinfo {author} {\bibfnamefont {M.}~\bibnamefont {Nitta}},\ and\ \bibinfo {author} {\bibfnamefont {N.}~\bibnamefont {Sakai}},\ }\href {https://doi.org/10.1007/JHEP03(2021)082} {\bibfield  {journal} {\bibinfo  {journal} {JHEP}\ }\textbf {\bibinfo {volume} {03}},\ \bibinfo {pages} {082}},\ \Eprint {https://arxiv.org/abs/2010.16080} {arXiv:2010.16080 [hep-th]} \BibitemShut {NoStop}%
\bibitem [{\citenamefont {Yu}(2023)}]{Yu:2023cic}%
  \BibitemOpen
  \bibfield  {author} {\bibinfo {author} {\bibfnamefont {C.}~\bibnamefont {Yu}},\ }\href {https://doi.org/10.1103/PhysRevD.108.116009} {\bibfield  {journal} {\bibinfo  {journal} {Phys. Rev. D}\ }\textbf {\bibinfo {volume} {108}},\ \bibinfo {pages} {116009} (\bibinfo {year} {2023})},\ \Eprint {https://arxiv.org/abs/2309.03570} {arXiv:2309.03570 [hep-ph]} \BibitemShut {NoStop}%
\bibitem [{\citenamefont {Aleksandrov}\ \emph {et~al.}(2024)\citenamefont {Aleksandrov}, \citenamefont {Kudlis},\ and\ \citenamefont {Klochai}}]{Aleksandrov:2024rsz}%
  \BibitemOpen
  \bibfield  {author} {\bibinfo {author} {\bibfnamefont {I.~A.}\ \bibnamefont {Aleksandrov}}, \bibinfo {author} {\bibfnamefont {A.}~\bibnamefont {Kudlis}},\ and\ \bibinfo {author} {\bibfnamefont {A.~I.}\ \bibnamefont {Klochai}},\ }\href {https://doi.org/10.1103/PhysRevResearch.6.043009} {\bibfield  {journal} {\bibinfo  {journal} {Phys. Rev. Res.}\ }\textbf {\bibinfo {volume} {6}},\ \bibinfo {pages} {043009} (\bibinfo {year} {2024})},\ \Eprint {https://arxiv.org/abs/2403.17204} {arXiv:2403.17204 [hep-ph]} \BibitemShut {NoStop}%
\bibitem [{\citenamefont {Yoon}\ \emph {et~al.}(2021)\citenamefont {Yoon} \emph {et~al.}}]{Yoon:2021ony}%
  \BibitemOpen
  \bibfield  {author} {\bibinfo {author} {\bibfnamefont {J.~W.}\ \bibnamefont {Yoon}} \emph {et~al.},\ }\href {https://doi.org/10.1364/OPTICA.420520} {\bibfield  {journal} {\bibinfo  {journal} {Optica}\ }\textbf {\bibinfo {volume} {8}},\ \bibinfo {pages} {630} (\bibinfo {year} {2021})}\BibitemShut {NoStop}%
\bibitem [{\citenamefont {Kohlf\"urst}(2024)}]{Kohlfurst:2022edl}%
  \BibitemOpen
  \bibfield  {author} {\bibinfo {author} {\bibfnamefont {C.}~\bibnamefont {Kohlf\"urst}},\ }\href {https://doi.org/10.1103/PhysRevD.110.L111903} {\bibfield  {journal} {\bibinfo  {journal} {Phys. Rev. D}\ }\textbf {\bibinfo {volume} {110}},\ \bibinfo {pages} {L111903} (\bibinfo {year} {2024})},\ \Eprint {https://arxiv.org/abs/2212.03180} {arXiv:2212.03180 [hep-ph]} \BibitemShut {NoStop}%
\bibitem [{\citenamefont {Hu}\ \emph {et~al.}(2023)\citenamefont {Hu}, \citenamefont {Amat}, \citenamefont {Wang}, \citenamefont {Sawut}, \citenamefont {Fan},\ and\ \citenamefont {Xie}}]{Hu:2023pmz}%
  \BibitemOpen
  \bibfield  {author} {\bibinfo {author} {\bibfnamefont {L.-N.}\ \bibnamefont {Hu}}, \bibinfo {author} {\bibfnamefont {O.}~\bibnamefont {Amat}}, \bibinfo {author} {\bibfnamefont {L.}~\bibnamefont {Wang}}, \bibinfo {author} {\bibfnamefont {A.}~\bibnamefont {Sawut}}, \bibinfo {author} {\bibfnamefont {H.-H.}\ \bibnamefont {Fan}},\ and\ \bibinfo {author} {\bibfnamefont {B.~S.}\ \bibnamefont {Xie}},\ }\href {https://doi.org/10.1103/PhysRevD.107.116010} {\bibfield  {journal} {\bibinfo  {journal} {Phys. Rev. D}\ }\textbf {\bibinfo {volume} {107}},\ \bibinfo {pages} {116010} (\bibinfo {year} {2023})},\ \bibinfo {note} {[Erratum: Phys.Rev.D 108, 039905 (2023)]},\ \Eprint {https://arxiv.org/abs/2303.15781} {arXiv:2303.15781 [hep-ph]} \BibitemShut {NoStop}%
\bibitem [{\citenamefont {Schutzhold}\ \emph {et~al.}(2008)\citenamefont {Schutzhold}, \citenamefont {Gies},\ and\ \citenamefont {Dunne}}]{Schutzhold:2008pz}%
  \BibitemOpen
  \bibfield  {author} {\bibinfo {author} {\bibfnamefont {R.}~\bibnamefont {Schutzhold}}, \bibinfo {author} {\bibfnamefont {H.}~\bibnamefont {Gies}},\ and\ \bibinfo {author} {\bibfnamefont {G.}~\bibnamefont {Dunne}},\ }\href {https://doi.org/10.1103/PhysRevLett.101.130404} {\bibfield  {journal} {\bibinfo  {journal} {Phys. Rev. Lett.}\ }\textbf {\bibinfo {volume} {101}},\ \bibinfo {pages} {130404} (\bibinfo {year} {2008})},\ \Eprint {https://arxiv.org/abs/0807.0754} {arXiv:0807.0754 [hep-th]} \BibitemShut {NoStop}%
\bibitem [{\citenamefont {Dunne}\ \emph {et~al.}(2009)\citenamefont {Dunne}, \citenamefont {Gies},\ and\ \citenamefont {Schutzhold}}]{Dunne:2009gi}%
  \BibitemOpen
  \bibfield  {author} {\bibinfo {author} {\bibfnamefont {G.~V.}\ \bibnamefont {Dunne}}, \bibinfo {author} {\bibfnamefont {H.}~\bibnamefont {Gies}},\ and\ \bibinfo {author} {\bibfnamefont {R.}~\bibnamefont {Schutzhold}},\ }\href {https://doi.org/10.1103/PhysRevD.80.111301} {\bibfield  {journal} {\bibinfo  {journal} {Phys. Rev. D}\ }\textbf {\bibinfo {volume} {80}},\ \bibinfo {pages} {111301} (\bibinfo {year} {2009})},\ \Eprint {https://arxiv.org/abs/0908.0948} {arXiv:0908.0948 [hep-ph]} \BibitemShut {NoStop}%
\bibitem [{\citenamefont {Torgrimsson}\ \emph {et~al.}(2016)\citenamefont {Torgrimsson}, \citenamefont {Oertel},\ and\ \citenamefont {Sch\"utzhold}}]{Torgrimsson:2016ant}%
  \BibitemOpen
  \bibfield  {author} {\bibinfo {author} {\bibfnamefont {G.}~\bibnamefont {Torgrimsson}}, \bibinfo {author} {\bibfnamefont {J.}~\bibnamefont {Oertel}},\ and\ \bibinfo {author} {\bibfnamefont {R.}~\bibnamefont {Sch\"utzhold}},\ }\href {https://doi.org/10.1103/PhysRevD.94.065035} {\bibfield  {journal} {\bibinfo  {journal} {Phys. Rev. D}\ }\textbf {\bibinfo {volume} {94}},\ \bibinfo {pages} {065035} (\bibinfo {year} {2016})},\ \Eprint {https://arxiv.org/abs/1607.02448} {arXiv:1607.02448 [hep-th]} \BibitemShut {NoStop}%
\bibitem [{\citenamefont {Torgrimsson}\ \emph {et~al.}(2017)\citenamefont {Torgrimsson}, \citenamefont {Schneider}, \citenamefont {Oertel},\ and\ \citenamefont {Sch\"utzhold}}]{Torgrimsson:2017pzs}%
  \BibitemOpen
  \bibfield  {author} {\bibinfo {author} {\bibfnamefont {G.}~\bibnamefont {Torgrimsson}}, \bibinfo {author} {\bibfnamefont {C.}~\bibnamefont {Schneider}}, \bibinfo {author} {\bibfnamefont {J.}~\bibnamefont {Oertel}},\ and\ \bibinfo {author} {\bibfnamefont {R.}~\bibnamefont {Sch\"utzhold}},\ }\href {https://doi.org/10.1007/JHEP06(2017)043} {\bibfield  {journal} {\bibinfo  {journal} {JHEP}\ }\textbf {\bibinfo {volume} {06}},\ \bibinfo {pages} {043}},\ \Eprint {https://arxiv.org/abs/1703.09203} {arXiv:1703.09203 [hep-th]} \BibitemShut {NoStop}%
\bibitem [{\citenamefont {W\"ollert}\ \emph {et~al.}(2015)\citenamefont {W\"ollert}, \citenamefont {Klaiber}, \citenamefont {Bauke},\ and\ \citenamefont {Keitel}}]{Wollert:2014epy}%
  \BibitemOpen
  \bibfield  {author} {\bibinfo {author} {\bibfnamefont {A.}~\bibnamefont {W\"ollert}}, \bibinfo {author} {\bibfnamefont {M.}~\bibnamefont {Klaiber}}, \bibinfo {author} {\bibfnamefont {H.}~\bibnamefont {Bauke}},\ and\ \bibinfo {author} {\bibfnamefont {C.~H.}\ \bibnamefont {Keitel}},\ }\href {https://doi.org/10.1103/PhysRevD.91.065022} {\bibfield  {journal} {\bibinfo  {journal} {Phys. Rev. D}\ }\textbf {\bibinfo {volume} {91}},\ \bibinfo {pages} {065022} (\bibinfo {year} {2015})},\ \Eprint {https://arxiv.org/abs/1410.2401} {arXiv:1410.2401 [quant-ph]} \BibitemShut {NoStop}%
\bibitem [{\citenamefont {Taya}(2020)}]{Taya:2020bcd}%
  \BibitemOpen
  \bibfield  {author} {\bibinfo {author} {\bibfnamefont {H.}~\bibnamefont {Taya}},\ }\href {https://doi.org/10.1103/PhysRevResearch.2.023257} {\bibfield  {journal} {\bibinfo  {journal} {Phys. Rev. Res.}\ }\textbf {\bibinfo {volume} {2}},\ \bibinfo {pages} {023257} (\bibinfo {year} {2020})},\ \Eprint {https://arxiv.org/abs/2003.08948} {arXiv:2003.08948 [hep-ph]} \BibitemShut {NoStop}%
\bibitem [{\citenamefont {Ababekri}\ \emph {et~al.}(2019)\citenamefont {Ababekri}, \citenamefont {Xie},\ and\ \citenamefont {Zhang}}]{Ababekri:2019dkl}%
  \BibitemOpen
  \bibfield  {author} {\bibinfo {author} {\bibfnamefont {M.}~\bibnamefont {Ababekri}}, \bibinfo {author} {\bibfnamefont {B.-S.}\ \bibnamefont {Xie}},\ and\ \bibinfo {author} {\bibfnamefont {J.}~\bibnamefont {Zhang}},\ }\href {https://doi.org/10.1103/PhysRevD.100.016003} {\bibfield  {journal} {\bibinfo  {journal} {Phys. Rev. D}\ }\textbf {\bibinfo {volume} {100}},\ \bibinfo {pages} {016003} (\bibinfo {year} {2019})},\ \Eprint {https://arxiv.org/abs/1905.01629} {arXiv:1905.01629 [hep-ph]} \BibitemShut {NoStop}%
\bibitem [{\citenamefont {Fedotov}\ and\ \citenamefont {Mironov}(2013)}]{Fedotov:2013uja}%
  \BibitemOpen
  \bibfield  {author} {\bibinfo {author} {\bibfnamefont {A.~M.}\ \bibnamefont {Fedotov}}\ and\ \bibinfo {author} {\bibfnamefont {A.~A.}\ \bibnamefont {Mironov}},\ }\href {https://doi.org/10.1103/PhysRevA.88.062110} {\bibfield  {journal} {\bibinfo  {journal} {Phys. Rev. A}\ }\textbf {\bibinfo {volume} {88}},\ \bibinfo {pages} {062110} (\bibinfo {year} {2013})},\ \Eprint {https://arxiv.org/abs/1310.7258} {arXiv:1310.7258 [hep-ph]} \BibitemShut {NoStop}%
\bibitem [{\citenamefont {Broers}\ and\ \citenamefont {Mathey}(2022)}]{PhysRevResearch.4.013057}%
  \BibitemOpen
  \bibfield  {author} {\bibinfo {author} {\bibfnamefont {L.}~\bibnamefont {Broers}}\ and\ \bibinfo {author} {\bibfnamefont {L.}~\bibnamefont {Mathey}},\ }\href {https://doi.org/10.1103/PhysRevResearch.4.013057} {\bibfield  {journal} {\bibinfo  {journal} {Phys. Rev. Res.}\ }\textbf {\bibinfo {volume} {4}},\ \bibinfo {pages} {013057} (\bibinfo {year} {2022})}\BibitemShut {NoStop}%
\bibitem [{\citenamefont {Oka}\ and\ \citenamefont {Aoki}(2009)}]{PhysRevB.79.081406}%
  \BibitemOpen
  \bibfield  {author} {\bibinfo {author} {\bibfnamefont {T.}~\bibnamefont {Oka}}\ and\ \bibinfo {author} {\bibfnamefont {H.}~\bibnamefont {Aoki}},\ }\href {https://doi.org/10.1103/PhysRevB.79.081406} {\bibfield  {journal} {\bibinfo  {journal} {Phys. Rev. B}\ }\textbf {\bibinfo {volume} {79}},\ \bibinfo {pages} {081406} (\bibinfo {year} {2009})}\BibitemShut {NoStop}%
\bibitem [{\citenamefont {Sato}\ \emph {et~al.}(2019)\citenamefont {Sato}, \citenamefont {McIver}, \citenamefont {Nuske}, \citenamefont {Tang}, \citenamefont {Jotzu}, \citenamefont {Schulte}, \citenamefont {H\"ubener}, \citenamefont {De~Giovannini}, \citenamefont {Mathey}, \citenamefont {Sentef}, \citenamefont {Cavalleri},\ and\ \citenamefont {Rubio}}]{PhysRevB.99.214302}%
  \BibitemOpen
  \bibfield  {author} {\bibinfo {author} {\bibfnamefont {S.~A.}\ \bibnamefont {Sato}}, \bibinfo {author} {\bibfnamefont {J.~W.}\ \bibnamefont {McIver}}, \bibinfo {author} {\bibfnamefont {M.}~\bibnamefont {Nuske}}, \bibinfo {author} {\bibfnamefont {P.}~\bibnamefont {Tang}}, \bibinfo {author} {\bibfnamefont {G.}~\bibnamefont {Jotzu}}, \bibinfo {author} {\bibfnamefont {B.}~\bibnamefont {Schulte}}, \bibinfo {author} {\bibfnamefont {H.}~\bibnamefont {H\"ubener}}, \bibinfo {author} {\bibfnamefont {U.}~\bibnamefont {De~Giovannini}}, \bibinfo {author} {\bibfnamefont {L.}~\bibnamefont {Mathey}}, \bibinfo {author} {\bibfnamefont {M.~A.}\ \bibnamefont {Sentef}}, \bibinfo {author} {\bibfnamefont {A.}~\bibnamefont {Cavalleri}},\ and\ \bibinfo {author} {\bibfnamefont {A.}~\bibnamefont {Rubio}},\ }\href {https://doi.org/10.1103/PhysRevB.99.214302} {\bibfield  {journal} {\bibinfo  {journal} {Phys. Rev. B}\ }\textbf {\bibinfo {volume} {99}},\ \bibinfo {pages} {214302} (\bibinfo {year} {2019})}\BibitemShut {NoStop}%
\bibitem [{\citenamefont {Luo}(2024)}]{Luo:2023abp}%
  \BibitemOpen
  \bibfield  {author} {\bibinfo {author} {\bibfnamefont {M.}~\bibnamefont {Luo}},\ }\href {https://doi.org/10.1103/PhysRevB.109.155401} {\bibfield  {journal} {\bibinfo  {journal} {Phys. Rev. B}\ }\textbf {\bibinfo {volume} {109}},\ \bibinfo {pages} {155401} (\bibinfo {year} {2024})},\ \Eprint {https://arxiv.org/abs/2311.12350} {arXiv:2311.12350 [cond-mat.mes-hall]} \BibitemShut {NoStop}%
\bibitem [{\citenamefont {Shi}\ \emph {et~al.}(2024)\citenamefont {Shi}, \citenamefont {Zhang},\ and\ \citenamefont {Zhang}}]{Shi:2024rem}%
  \BibitemOpen
  \bibfield  {author} {\bibinfo {author} {\bibfnamefont {K.}~\bibnamefont {Shi}}, \bibinfo {author} {\bibfnamefont {X.}~\bibnamefont {Zhang}},\ and\ \bibinfo {author} {\bibfnamefont {W.}~\bibnamefont {Zhang}},\ }\href {https://doi.org/10.1103/PhysRevA.109.013324} {\bibfield  {journal} {\bibinfo  {journal} {Phys. Rev. A}\ }\textbf {\bibinfo {volume} {109}},\ \bibinfo {pages} {013324} (\bibinfo {year} {2024})},\ \Eprint {https://arxiv.org/abs/2401.01250} {arXiv:2401.01250 [cond-mat.quant-gas]} \BibitemShut {NoStop}%
\bibitem [{\citenamefont {Mumford}\ \emph {et~al.}(2024)\citenamefont {Mumford}, \citenamefont {Kamp},\ and\ \citenamefont {O'Dell}}]{Mumford:2024yem}%
  \BibitemOpen
  \bibfield  {author} {\bibinfo {author} {\bibfnamefont {J.}~\bibnamefont {Mumford}}, \bibinfo {author} {\bibfnamefont {D.}~\bibnamefont {Kamp}},\ and\ \bibinfo {author} {\bibfnamefont {D.~H.~J.}\ \bibnamefont {O'Dell}},\ }\href {https://doi.org/10.1103/PhysRevA.110.043310} {\bibfield  {journal} {\bibinfo  {journal} {Phys. Rev. A}\ }\textbf {\bibinfo {volume} {110}},\ \bibinfo {pages} {043310} (\bibinfo {year} {2024})},\ \Eprint {https://arxiv.org/abs/2404.00533} {arXiv:2404.00533 [cond-mat.quant-gas]} \BibitemShut {NoStop}%
\bibitem [{\citenamefont {Wang}\ \emph {et~al.}(2024)\citenamefont {Wang}, \citenamefont {Zhang}, \citenamefont {Tang},\ and\ \citenamefont {Wang}}]{Wang:2024bbb}%
  \BibitemOpen
  \bibfield  {author} {\bibinfo {author} {\bibfnamefont {W.}~\bibnamefont {Wang}}, \bibinfo {author} {\bibfnamefont {Z.}~\bibnamefont {Zhang}}, \bibinfo {author} {\bibfnamefont {G.-X.}\ \bibnamefont {Tang}},\ and\ \bibinfo {author} {\bibfnamefont {T.}~\bibnamefont {Wang}},\ }\href {https://doi.org/10.1103/PhysRevA.110.023308} {\bibfield  {journal} {\bibinfo  {journal} {Phys. Rev. A}\ }\textbf {\bibinfo {volume} {110}},\ \bibinfo {pages} {023308} (\bibinfo {year} {2024})},\ \Eprint {https://arxiv.org/abs/2404.12682} {arXiv:2404.12682 [cond-mat.quant-gas]} \BibitemShut {NoStop}%
\bibitem [{\citenamefont {Chakrabarty}\ and\ \citenamefont {Datta}(2024)}]{Chakrabarty:2024hpw}%
  \BibitemOpen
  \bibfield  {author} {\bibinfo {author} {\bibfnamefont {A.}~\bibnamefont {Chakrabarty}}\ and\ \bibinfo {author} {\bibfnamefont {S.}~\bibnamefont {Datta}},\ }\href {https://doi.org/10.1103/PhysRevB.110.064314} {\bibfield  {journal} {\bibinfo  {journal} {Phys. Rev. B}\ }\textbf {\bibinfo {volume} {110}},\ \bibinfo {pages} {064314} (\bibinfo {year} {2024})},\ \Eprint {https://arxiv.org/abs/2404.12297} {arXiv:2404.12297 [cond-mat.dis-nn]} \BibitemShut {NoStop}%
\bibitem [{\citenamefont {Schnell}\ \emph {et~al.}(2023)\citenamefont {Schnell}, \citenamefont {Wu}, \citenamefont {Widera},\ and\ \citenamefont {Eckardt}}]{Schnell:2022nhn}%
  \BibitemOpen
  \bibfield  {author} {\bibinfo {author} {\bibfnamefont {A.}~\bibnamefont {Schnell}}, \bibinfo {author} {\bibfnamefont {L.-N.}\ \bibnamefont {Wu}}, \bibinfo {author} {\bibfnamefont {A.}~\bibnamefont {Widera}},\ and\ \bibinfo {author} {\bibfnamefont {A.}~\bibnamefont {Eckardt}},\ }\href {https://doi.org/10.1103/PhysRevA.107.L021301} {\bibfield  {journal} {\bibinfo  {journal} {Phys. Rev. A}\ }\textbf {\bibinfo {volume} {107}},\ \bibinfo {pages} {L021301} (\bibinfo {year} {2023})},\ \Eprint {https://arxiv.org/abs/2204.07147} {arXiv:2204.07147 [cond-mat.quant-gas]} \BibitemShut {NoStop}%
\bibitem [{\citenamefont {Blanes}\ \emph {et~al.}(2009)\citenamefont {Blanes}, \citenamefont {Casas}, \citenamefont {Oteo},\ and\ \citenamefont {Ros}}]{Blanes:2008xlr}%
  \BibitemOpen
  \bibfield  {author} {\bibinfo {author} {\bibfnamefont {S.}~\bibnamefont {Blanes}}, \bibinfo {author} {\bibfnamefont {F.}~\bibnamefont {Casas}}, \bibinfo {author} {\bibfnamefont {J.~A.}\ \bibnamefont {Oteo}},\ and\ \bibinfo {author} {\bibfnamefont {J.}~\bibnamefont {Ros}},\ }\href {https://doi.org/10.1016/j.physrep.2008.11.001} {\bibfield  {journal} {\bibinfo  {journal} {Phys. Rept.}\ }\textbf {\bibinfo {volume} {470}},\ \bibinfo {pages} {151} (\bibinfo {year} {2009})}\BibitemShut {NoStop}%
\bibitem [{\citenamefont {Mananga}\ and\ \citenamefont {Charpentier}(2016)}]{Managa2016}%
  \BibitemOpen
  \bibfield  {author} {\bibinfo {author} {\bibfnamefont {E.~S.}\ \bibnamefont {Mananga}}\ and\ \bibinfo {author} {\bibfnamefont {T.}~\bibnamefont {Charpentier}},\ }\href {https://doi.org/https://doi.org/10.1016/j.physrep.2015.10.005} {\bibfield  {journal} {\bibinfo  {journal} {Physics Reports}\ }\textbf {\bibinfo {volume} {609}},\ \bibinfo {pages} {1} (\bibinfo {year} {2016})}\BibitemShut {NoStop}%
\bibitem [{\citenamefont {Yamada}\ and\ \citenamefont {Yamamoto}(2021)}]{Yamada:2021jhy}%
  \BibitemOpen
  \bibfield  {author} {\bibinfo {author} {\bibfnamefont {A.}~\bibnamefont {Yamada}}\ and\ \bibinfo {author} {\bibfnamefont {N.}~\bibnamefont {Yamamoto}},\ }\href {https://doi.org/10.1103/PhysRevD.104.054041} {\bibfield  {journal} {\bibinfo  {journal} {Phys. Rev. D}\ }\textbf {\bibinfo {volume} {104}},\ \bibinfo {pages} {054041} (\bibinfo {year} {2021})},\ \Eprint {https://arxiv.org/abs/2107.07074} {arXiv:2107.07074 [hep-ph]} \BibitemShut {NoStop}%
\bibitem [{\citenamefont {Fukushima}\ \emph {et~al.}(2023)\citenamefont {Fukushima}, \citenamefont {Hidaka}, \citenamefont {Shimazaki},\ and\ \citenamefont {Taya}}]{Fukushima:2023obj}%
  \BibitemOpen
  \bibfield  {author} {\bibinfo {author} {\bibfnamefont {K.}~\bibnamefont {Fukushima}}, \bibinfo {author} {\bibfnamefont {Y.}~\bibnamefont {Hidaka}}, \bibinfo {author} {\bibfnamefont {T.}~\bibnamefont {Shimazaki}},\ and\ \bibinfo {author} {\bibfnamefont {H.}~\bibnamefont {Taya}},\ }\href {https://doi.org/10.1016/j.aop.2023.169494} {\bibfield  {journal} {\bibinfo  {journal} {Annals Phys.}\ }\textbf {\bibinfo {volume} {458}},\ \bibinfo {pages} {169494} (\bibinfo {year} {2023})},\ \Eprint {https://arxiv.org/abs/2305.11432} {arXiv:2305.11432 [hep-ph]} \BibitemShut {NoStop}%
\bibitem [{\citenamefont {Marin~Bukov}\ and\ \citenamefont {Polkovnikov}(2015)}]{Bukov04032015}%
  \BibitemOpen
  \bibfield  {author} {\bibinfo {author} {\bibfnamefont {L.~D.}\ \bibnamefont {Marin~Bukov}}\ and\ \bibinfo {author} {\bibfnamefont {A.}~\bibnamefont {Polkovnikov}},\ }\href {https://doi.org/10.1080/00018732.2015.1055918} {\bibfield  {journal} {\bibinfo  {journal} {Advances in Physics}\ }\textbf {\bibinfo {volume} {64}},\ \bibinfo {pages} {139} (\bibinfo {year} {2015})},\ \Eprint {https://arxiv.org/abs/https://doi.org/10.1080/00018732.2015.1055918} {https://doi.org/10.1080/00018732.2015.1055918} \BibitemShut {NoStop}%
\bibitem [{\citenamefont {Magnus}(1954)}]{Magnus:1954zz}%
  \BibitemOpen
  \bibfield  {author} {\bibinfo {author} {\bibfnamefont {W.}~\bibnamefont {Magnus}},\ }\href {https://doi.org/10.1002/cpa.3160070404} {\bibfield  {journal} {\bibinfo  {journal} {Commun. Pure Appl. Math.}\ }\textbf {\bibinfo {volume} {7}},\ \bibinfo {pages} {649} (\bibinfo {year} {1954})}\BibitemShut {NoStop}%
\bibitem [{\citenamefont {Rahav}\ \emph {et~al.}(2003)\citenamefont {Rahav}, \citenamefont {Gilary},\ and\ \citenamefont {Fishman}}]{PhysRevA.68.013820}%
  \BibitemOpen
  \bibfield  {author} {\bibinfo {author} {\bibfnamefont {S.}~\bibnamefont {Rahav}}, \bibinfo {author} {\bibfnamefont {I.}~\bibnamefont {Gilary}},\ and\ \bibinfo {author} {\bibfnamefont {S.}~\bibnamefont {Fishman}},\ }\href {https://doi.org/10.1103/PhysRevA.68.013820} {\bibfield  {journal} {\bibinfo  {journal} {Phys. Rev. A}\ }\textbf {\bibinfo {volume} {68}},\ \bibinfo {pages} {013820} (\bibinfo {year} {2003})}\BibitemShut {NoStop}%
\bibitem [{\citenamefont {Oka}\ and\ \citenamefont {Kitamura}(2019)}]{annurev:/content/journals/10.1146/annurev-conmatphys-031218-013423}%
  \BibitemOpen
  \bibfield  {author} {\bibinfo {author} {\bibfnamefont {T.}~\bibnamefont {Oka}}\ and\ \bibinfo {author} {\bibfnamefont {S.}~\bibnamefont {Kitamura}},\ }\href {https://doi.org/https://doi.org/10.1146/annurev-conmatphys-031218-013423} {\bibfield  {journal} {\bibinfo  {journal} {Annual Review of Condensed Matter Physics}\ }\textbf {\bibinfo {volume} {10}},\ \bibinfo {pages} {387} (\bibinfo {year} {2019})}\BibitemShut {NoStop}%
\bibitem [{\citenamefont {Goldman}\ and\ \citenamefont {Dalibard}(2014)}]{Goldman:2014xja}%
  \BibitemOpen
  \bibfield  {author} {\bibinfo {author} {\bibfnamefont {N.}~\bibnamefont {Goldman}}\ and\ \bibinfo {author} {\bibfnamefont {J.}~\bibnamefont {Dalibard}},\ }\href {https://doi.org/10.1103/PhysRevX.4.031027} {\bibfield  {journal} {\bibinfo  {journal} {Phys. Rev. X}\ }\textbf {\bibinfo {volume} {4}},\ \bibinfo {pages} {031027} (\bibinfo {year} {2014})},\ \bibinfo {note} {[Erratum: Phys.Rev.X 5, 029902 (2015)]},\ \Eprint {https://arxiv.org/abs/1404.4373} {arXiv:1404.4373 [cond-mat.quant-gas]} \BibitemShut {NoStop}%
\bibitem [{\citenamefont {Aidelsburger}\ \emph {et~al.}(2018)\citenamefont {Aidelsburger}, \citenamefont {Nascimbene},\ and\ \citenamefont {Goldman}}]{Aidelsburger:2017qlh}%
  \BibitemOpen
  \bibfield  {author} {\bibinfo {author} {\bibfnamefont {M.}~\bibnamefont {Aidelsburger}}, \bibinfo {author} {\bibfnamefont {S.}~\bibnamefont {Nascimbene}},\ and\ \bibinfo {author} {\bibfnamefont {N.}~\bibnamefont {Goldman}},\ }\href {https://doi.org/10.1016/j.crhy.2018.03.002} {\bibfield  {journal} {\bibinfo  {journal} {Comptes Rendus Physique}\ }\textbf {\bibinfo {volume} {19}},\ \bibinfo {pages} {394} (\bibinfo {year} {2018})},\ \Eprint {https://arxiv.org/abs/1710.00851} {arXiv:1710.00851 [cond-mat.mes-hall]} \BibitemShut {NoStop}%
\bibitem [{\citenamefont {Eckardt}\ and\ \citenamefont {Anisimovas}(2015)}]{Eckardt:2015mtt}%
  \BibitemOpen
  \bibfield  {author} {\bibinfo {author} {\bibfnamefont {A.}~\bibnamefont {Eckardt}}\ and\ \bibinfo {author} {\bibfnamefont {E.}~\bibnamefont {Anisimovas}},\ }\href {https://doi.org/10.1088/1367-2630/17/9/093039} {\bibfield  {journal} {\bibinfo  {journal} {New J. Phys.}\ }\textbf {\bibinfo {volume} {17}},\ \bibinfo {pages} {093039} (\bibinfo {year} {2015})}\BibitemShut {NoStop}%
\bibitem [{\citenamefont {Uhrig}\ \emph {et~al.}(2019)\citenamefont {Uhrig}, \citenamefont {Kalthoff},\ and\ \citenamefont {Freericks}}]{PhysRevLett.122.130604}%
  \BibitemOpen
  \bibfield  {author} {\bibinfo {author} {\bibfnamefont {G.~S.}\ \bibnamefont {Uhrig}}, \bibinfo {author} {\bibfnamefont {M.~H.}\ \bibnamefont {Kalthoff}},\ and\ \bibinfo {author} {\bibfnamefont {J.~K.}\ \bibnamefont {Freericks}},\ }\href {https://doi.org/10.1103/PhysRevLett.122.130604} {\bibfield  {journal} {\bibinfo  {journal} {Phys. Rev. Lett.}\ }\textbf {\bibinfo {volume} {122}},\ \bibinfo {pages} {130604} (\bibinfo {year} {2019})}\BibitemShut {NoStop}%
\bibitem [{\citenamefont {Rodriguez-Vega}\ \emph {et~al.}(2021)\citenamefont {Rodriguez-Vega}, \citenamefont {Vogl},\ and\ \citenamefont {Fiete}}]{RODRIGUEZVEGA2021168434}%
  \BibitemOpen
  \bibfield  {author} {\bibinfo {author} {\bibfnamefont {M.}~\bibnamefont {Rodriguez-Vega}}, \bibinfo {author} {\bibfnamefont {M.}~\bibnamefont {Vogl}},\ and\ \bibinfo {author} {\bibfnamefont {G.~A.}\ \bibnamefont {Fiete}},\ }\href {https://doi.org/https://doi.org/10.1016/j.aop.2021.168434} {\bibfield  {journal} {\bibinfo  {journal} {Annals of Physics}\ }\textbf {\bibinfo {volume} {435}},\ \bibinfo {pages} {168434} (\bibinfo {year} {2021})}\BibitemShut {NoStop}%
\bibitem [{\citenamefont {Ilan}\ \emph {et~al.}(2019)\citenamefont {Ilan}, \citenamefont {Grushin},\ and\ \citenamefont {Pikulin}}]{Ilan:2019lqk}%
  \BibitemOpen
  \bibfield  {author} {\bibinfo {author} {\bibfnamefont {R.}~\bibnamefont {Ilan}}, \bibinfo {author} {\bibfnamefont {A.~G.}\ \bibnamefont {Grushin}},\ and\ \bibinfo {author} {\bibfnamefont {D.~I.}\ \bibnamefont {Pikulin}},\ }\href {https://doi.org/10.1038/s42254-019-0121-8} {\bibfield  {journal} {\bibinfo  {journal} {Nature Rev. Phys.}\ }\textbf {\bibinfo {volume} {2}},\ \bibinfo {pages} {29} (\bibinfo {year} {2019})},\ \Eprint {https://arxiv.org/abs/1903.11088} {arXiv:1903.11088 [cond-mat.mes-hall]} \BibitemShut {NoStop}%
\bibitem [{\citenamefont {Jamotte}\ \emph {et~al.}(2022)\citenamefont {Jamotte}, \citenamefont {Goldman},\ and\ \citenamefont {Di~Liberto}}]{Jamotte:2021mvq}%
  \BibitemOpen
  \bibfield  {author} {\bibinfo {author} {\bibfnamefont {M.}~\bibnamefont {Jamotte}}, \bibinfo {author} {\bibfnamefont {N.}~\bibnamefont {Goldman}},\ and\ \bibinfo {author} {\bibfnamefont {M.}~\bibnamefont {Di~Liberto}},\ }\href {https://doi.org/10.1038/s42005-022-00802-9} {\bibfield  {journal} {\bibinfo  {journal} {Commun. Phys.}\ }\textbf {\bibinfo {volume} {5}},\ \bibinfo {pages} {30} (\bibinfo {year} {2022})},\ \Eprint {https://arxiv.org/abs/2104.13394} {arXiv:2104.13394 [cond-mat.quant-gas]} \BibitemShut {NoStop}%
\bibitem [{\citenamefont {Chernodub}\ \emph {et~al.}(2022)\citenamefont {Chernodub}, \citenamefont {Ferreiros}, \citenamefont {Grushin}, \citenamefont {Landsteiner},\ and\ \citenamefont {Vozmediano}}]{Chernodub:2021nff}%
  \BibitemOpen
  \bibfield  {author} {\bibinfo {author} {\bibfnamefont {M.~N.}\ \bibnamefont {Chernodub}}, \bibinfo {author} {\bibfnamefont {Y.}~\bibnamefont {Ferreiros}}, \bibinfo {author} {\bibfnamefont {A.~G.}\ \bibnamefont {Grushin}}, \bibinfo {author} {\bibfnamefont {K.}~\bibnamefont {Landsteiner}},\ and\ \bibinfo {author} {\bibfnamefont {M.~A.~H.}\ \bibnamefont {Vozmediano}},\ }\href {https://doi.org/10.1016/j.physrep.2022.06.002} {\bibfield  {journal} {\bibinfo  {journal} {Phys. Rept.}\ }\textbf {\bibinfo {volume} {977}},\ \bibinfo {pages} {1} (\bibinfo {year} {2022})},\ \Eprint {https://arxiv.org/abs/2110.05471} {arXiv:2110.05471 [cond-mat.mes-hall]} \BibitemShut {NoStop}%
\bibitem [{\citenamefont {Cortijo}\ \emph {et~al.}(2015)\citenamefont {Cortijo}, \citenamefont {Ferreiros}, \citenamefont {Landsteiner},\ and\ \citenamefont {Vozmediano}}]{Cortijo:2015hlt}%
  \BibitemOpen
  \bibfield  {author} {\bibinfo {author} {\bibfnamefont {A.}~\bibnamefont {Cortijo}}, \bibinfo {author} {\bibfnamefont {Y.}~\bibnamefont {Ferreiros}}, \bibinfo {author} {\bibfnamefont {K.}~\bibnamefont {Landsteiner}},\ and\ \bibinfo {author} {\bibfnamefont {M.~A.~H.}\ \bibnamefont {Vozmediano}},\ }\href {https://doi.org/10.1103/PhysRevLett.115.177202} {\bibfield  {journal} {\bibinfo  {journal} {Phys. Rev. Lett.}\ }\textbf {\bibinfo {volume} {115}},\ \bibinfo {pages} {177202} (\bibinfo {year} {2015})},\ \Eprint {https://arxiv.org/abs/1603.02674} {arXiv:1603.02674 [cond-mat.mes-hall]} \BibitemShut {NoStop}%
\bibitem [{\citenamefont {Arjona}\ \emph {et~al.}(2017)\citenamefont {Arjona}, \citenamefont {Castro},\ and\ \citenamefont {Vozmediano}}]{Arjona:2017qjc}%
  \BibitemOpen
  \bibfield  {author} {\bibinfo {author} {\bibfnamefont {V.}~\bibnamefont {Arjona}}, \bibinfo {author} {\bibfnamefont {E.~V.}\ \bibnamefont {Castro}},\ and\ \bibinfo {author} {\bibfnamefont {M.~A.~H.}\ \bibnamefont {Vozmediano}},\ }\href {https://doi.org/10.1103/PhysRevB.96.081110} {\bibfield  {journal} {\bibinfo  {journal} {Phys. Rev. B}\ }\textbf {\bibinfo {volume} {96}},\ \bibinfo {pages} {081110} (\bibinfo {year} {2017})},\ \Eprint {https://arxiv.org/abs/1703.05399} {arXiv:1703.05399 [cond-mat.mes-hall]} \BibitemShut {NoStop}%
\bibitem [{\citenamefont {Gorbar}\ \emph {et~al.}(2017)\citenamefont {Gorbar}, \citenamefont {Miransky}, \citenamefont {Shovkovy},\ and\ \citenamefont {Sukhachov}}]{Gorbar:2016ygi}%
  \BibitemOpen
  \bibfield  {author} {\bibinfo {author} {\bibfnamefont {E.~V.}\ \bibnamefont {Gorbar}}, \bibinfo {author} {\bibfnamefont {V.~A.}\ \bibnamefont {Miransky}}, \bibinfo {author} {\bibfnamefont {I.~A.}\ \bibnamefont {Shovkovy}},\ and\ \bibinfo {author} {\bibfnamefont {P.~O.}\ \bibnamefont {Sukhachov}},\ }\href {https://doi.org/10.1103/PhysRevLett.118.127601} {\bibfield  {journal} {\bibinfo  {journal} {Phys. Rev. Lett.}\ }\textbf {\bibinfo {volume} {118}},\ \bibinfo {pages} {127601} (\bibinfo {year} {2017})},\ \Eprint {https://arxiv.org/abs/1610.07625} {arXiv:1610.07625 [cond-mat.str-el]} \BibitemShut {NoStop}%
\bibitem [{\citenamefont {Landsteiner}\ and\ \citenamefont {Liu}(2018)}]{Landsteiner:2017hye}%
  \BibitemOpen
  \bibfield  {author} {\bibinfo {author} {\bibfnamefont {K.}~\bibnamefont {Landsteiner}}\ and\ \bibinfo {author} {\bibfnamefont {Y.}~\bibnamefont {Liu}},\ }\href {https://doi.org/10.1016/j.physletb.2018.04.068} {\bibfield  {journal} {\bibinfo  {journal} {Phys. Lett. B}\ }\textbf {\bibinfo {volume} {783}},\ \bibinfo {pages} {446} (\bibinfo {year} {2018})},\ \Eprint {https://arxiv.org/abs/1703.01944} {arXiv:1703.01944 [cond-mat.mes-hall]} \BibitemShut {NoStop}%
\bibitem [{\citenamefont {Yamamoto}\ and\ \citenamefont {Yang}(2021)}]{Yamamoto:2021gts}%
  \BibitemOpen
  \bibfield  {author} {\bibinfo {author} {\bibfnamefont {N.}~\bibnamefont {Yamamoto}}\ and\ \bibinfo {author} {\bibfnamefont {D.-L.}\ \bibnamefont {Yang}},\ }\href {https://doi.org/10.1103/PhysRevD.103.125003} {\bibfield  {journal} {\bibinfo  {journal} {Phys. Rev. D}\ }\textbf {\bibinfo {volume} {103}},\ \bibinfo {pages} {125003} (\bibinfo {year} {2021})},\ \Eprint {https://arxiv.org/abs/2103.13208} {arXiv:2103.13208 [hep-th]} \BibitemShut {NoStop}%
\bibitem [{\citenamefont {Copinger}\ and\ \citenamefont {Pu}(2020)}]{Copinger:2020nyx}%
  \BibitemOpen
  \bibfield  {author} {\bibinfo {author} {\bibfnamefont {P.}~\bibnamefont {Copinger}}\ and\ \bibinfo {author} {\bibfnamefont {S.}~\bibnamefont {Pu}},\ }\href {https://doi.org/10.1142/S0217751X2030015X} {\bibfield  {journal} {\bibinfo  {journal} {Int. J. Mod. Phys. A}\ }\textbf {\bibinfo {volume} {35}},\ \bibinfo {pages} {2030015} (\bibinfo {year} {2020})},\ \Eprint {https://arxiv.org/abs/2008.03635} {arXiv:2008.03635 [hep-ph]} \BibitemShut {NoStop}%
\bibitem [{\citenamefont {Huang}\ and\ \citenamefont {Taya}(2019)}]{Huang:2019uhf}%
  \BibitemOpen
  \bibfield  {author} {\bibinfo {author} {\bibfnamefont {X.-G.}\ \bibnamefont {Huang}}\ and\ \bibinfo {author} {\bibfnamefont {H.}~\bibnamefont {Taya}},\ }\href {https://doi.org/10.1103/PhysRevD.100.016013} {\bibfield  {journal} {\bibinfo  {journal} {Phys. Rev. D}\ }\textbf {\bibinfo {volume} {100}},\ \bibinfo {pages} {016013} (\bibinfo {year} {2019})},\ \Eprint {https://arxiv.org/abs/1904.08200} {arXiv:1904.08200 [hep-ph]} \BibitemShut {NoStop}%
\bibitem [{\citenamefont {Aleksandrov}\ and\ \citenamefont {Kudlis}(2024)}]{Aleksandrov:2024cqh}%
  \BibitemOpen
  \bibfield  {author} {\bibinfo {author} {\bibfnamefont {I.~A.}\ \bibnamefont {Aleksandrov}}\ and\ \bibinfo {author} {\bibfnamefont {A.}~\bibnamefont {Kudlis}},\ }\href {https://doi.org/10.1103/PhysRevD.110.L011901} {\bibfield  {journal} {\bibinfo  {journal} {Phys. Rev. D}\ }\textbf {\bibinfo {volume} {110}},\ \bibinfo {pages} {L011901} (\bibinfo {year} {2024})},\ \Eprint {https://arxiv.org/abs/2404.02878} {arXiv:2404.02878 [hep-ph]} \BibitemShut {NoStop}%
\bibitem [{\citenamefont {Fukushima}\ \emph {et~al.}(2008)\citenamefont {Fukushima}, \citenamefont {Kharzeev},\ and\ \citenamefont {Warringa}}]{Fukushima:2008xe}%
  \BibitemOpen
  \bibfield  {author} {\bibinfo {author} {\bibfnamefont {K.}~\bibnamefont {Fukushima}}, \bibinfo {author} {\bibfnamefont {D.~E.}\ \bibnamefont {Kharzeev}},\ and\ \bibinfo {author} {\bibfnamefont {H.~J.}\ \bibnamefont {Warringa}},\ }\href {https://doi.org/10.1103/PhysRevD.78.074033} {\bibfield  {journal} {\bibinfo  {journal} {Phys. Rev. D}\ }\textbf {\bibinfo {volume} {78}},\ \bibinfo {pages} {074033} (\bibinfo {year} {2008})},\ \Eprint {https://arxiv.org/abs/0808.3382} {arXiv:0808.3382 [hep-ph]} \BibitemShut {NoStop}%
\bibitem [{\citenamefont {Domcke}\ \emph {et~al.}(2021)\citenamefont {Domcke}, \citenamefont {Ema},\ and\ \citenamefont {Mukaida}}]{Domcke:2021fee}%
  \BibitemOpen
  \bibfield  {author} {\bibinfo {author} {\bibfnamefont {V.}~\bibnamefont {Domcke}}, \bibinfo {author} {\bibfnamefont {Y.}~\bibnamefont {Ema}},\ and\ \bibinfo {author} {\bibfnamefont {K.}~\bibnamefont {Mukaida}},\ }\href {https://doi.org/10.1007/JHEP05(2021)001} {\bibfield  {journal} {\bibinfo  {journal} {JHEP}\ }\textbf {\bibinfo {volume} {05}},\ \bibinfo {pages} {001}},\ \Eprint {https://arxiv.org/abs/2101.05192} {arXiv:2101.05192 [hep-ph]} \BibitemShut {NoStop}%
\bibitem [{\citenamefont {Copinger}\ \emph {et~al.}(2023)\citenamefont {Copinger}, \citenamefont {Hattori},\ and\ \citenamefont {Yang}}]{Copinger:2022bwl}%
  \BibitemOpen
  \bibfield  {author} {\bibinfo {author} {\bibfnamefont {P.}~\bibnamefont {Copinger}}, \bibinfo {author} {\bibfnamefont {K.}~\bibnamefont {Hattori}},\ and\ \bibinfo {author} {\bibfnamefont {D.-L.}\ \bibnamefont {Yang}},\ }\href {https://doi.org/10.1103/PhysRevD.107.056016} {\bibfield  {journal} {\bibinfo  {journal} {Phys. Rev. D}\ }\textbf {\bibinfo {volume} {107}},\ \bibinfo {pages} {056016} (\bibinfo {year} {2023})},\ \Eprint {https://arxiv.org/abs/2208.12913} {arXiv:2208.12913 [hep-th]} \BibitemShut {NoStop}%
\bibitem [{\citenamefont {Furry}(1951)}]{Furry:1951bef}%
  \BibitemOpen
  \bibfield  {author} {\bibinfo {author} {\bibfnamefont {W.~H.}\ \bibnamefont {Furry}},\ }\href {https://doi.org/10.1103/PhysRev.81.115} {\bibfield  {journal} {\bibinfo  {journal} {Phys. Rev.}\ }\textbf {\bibinfo {volume} {81}},\ \bibinfo {pages} {115} (\bibinfo {year} {1951})}\BibitemShut {NoStop}%
\bibitem [{\citenamefont {Moortgat-Pick}(2009)}]{Moortgat-Pick:2009fyg}%
  \BibitemOpen
  \bibfield  {author} {\bibinfo {author} {\bibfnamefont {G.}~\bibnamefont {Moortgat-Pick}},\ }\href {https://doi.org/10.1088/1742-6596/198/1/012002} {\bibfield  {journal} {\bibinfo  {journal} {J. Phys. Conf. Ser.}\ }\textbf {\bibinfo {volume} {198}},\ \bibinfo {pages} {012002} (\bibinfo {year} {2009})}\BibitemShut {NoStop}%
\bibitem [{\citenamefont {Di~Piazza}(2015)}]{DiPiazza:2015xva}%
  \BibitemOpen
  \bibfield  {author} {\bibinfo {author} {\bibfnamefont {A.}~\bibnamefont {Di~Piazza}},\ }\href {https://doi.org/10.1103/PhysRevA.91.042118} {\bibfield  {journal} {\bibinfo  {journal} {Phys. Rev. A}\ }\textbf {\bibinfo {volume} {91}},\ \bibinfo {pages} {042118} (\bibinfo {year} {2015})},\ \Eprint {https://arxiv.org/abs/1501.06475} {arXiv:1501.06475 [hep-ph]} \BibitemShut {NoStop}%
\bibitem [{\citenamefont {Taya}\ \emph {et~al.}(2014)\citenamefont {Taya}, \citenamefont {Fujii},\ and\ \citenamefont {Itakura}}]{Taya:2014taa}%
  \BibitemOpen
  \bibfield  {author} {\bibinfo {author} {\bibfnamefont {H.}~\bibnamefont {Taya}}, \bibinfo {author} {\bibfnamefont {H.}~\bibnamefont {Fujii}},\ and\ \bibinfo {author} {\bibfnamefont {K.}~\bibnamefont {Itakura}},\ }\href {https://doi.org/10.1103/PhysRevD.90.014039} {\bibfield  {journal} {\bibinfo  {journal} {Phys. Rev. D}\ }\textbf {\bibinfo {volume} {90}},\ \bibinfo {pages} {014039} (\bibinfo {year} {2014})},\ \Eprint {https://arxiv.org/abs/1405.6182} {arXiv:1405.6182 [hep-ph]} \BibitemShut {NoStop}%
\bibitem [{\citenamefont {Fukushima}(2015)}]{Fukushima:2015tza}%
  \BibitemOpen
  \bibfield  {author} {\bibinfo {author} {\bibfnamefont {K.}~\bibnamefont {Fukushima}},\ }\href {https://doi.org/10.1103/PhysRevD.92.054009} {\bibfield  {journal} {\bibinfo  {journal} {Phys. Rev. D}\ }\textbf {\bibinfo {volume} {92}},\ \bibinfo {pages} {054009} (\bibinfo {year} {2015})},\ \Eprint {https://arxiv.org/abs/1501.01940} {arXiv:1501.01940 [hep-ph]} \BibitemShut {NoStop}%
\bibitem [{\citenamefont {Di~Piazza}(2014)}]{DiPiazza:2013vra}%
  \BibitemOpen
  \bibfield  {author} {\bibinfo {author} {\bibfnamefont {A.}~\bibnamefont {Di~Piazza}},\ }\href {https://doi.org/10.1103/PhysRevLett.113.040402} {\bibfield  {journal} {\bibinfo  {journal} {Phys. Rev. Lett.}\ }\textbf {\bibinfo {volume} {113}},\ \bibinfo {pages} {040402} (\bibinfo {year} {2014})},\ \Eprint {https://arxiv.org/abs/1310.7856} {arXiv:1310.7856 [hep-ph]} \BibitemShut {NoStop}%
\end{thebibliography}%

\end{document}